\documentclass[11pt]{article}

\usepackage[margin=1in]{geometry}
\usepackage{indentfirst}
\usepackage{times}
\usepackage{amsmath}
\usepackage{bm}
\usepackage{amssymb}
\usepackage{amsthm}
\usepackage{mathtools}
\usepackage{mleftright}
\usepackage{thmtools}
\usepackage{algorithm}
\usepackage[noend]{algpseudocode}
\usepackage[numbers]{natbib}
\usepackage[colorlinks=true,linkcolor=blue,citecolor=blue,urlcolor=blue]{hyperref}
\usepackage[nameinlink,capitalize]{cleveref}
\usepackage{enumitem}
\usepackage{mdframed}
\usepackage{tikz}
\usetikzlibrary{arrows.meta,calc,positioning}
\definecolor{corlinks}{RGB}{0,0,150}
\definecolor{cormenu}{RGB}{0,0,150}
\definecolor{corurl}{RGB}{0,0,150}

\hypersetup{
	colorlinks=true,
	urlcolor=corlinks,
	linkcolor=corlinks,
	menucolor=cormenu,
	citecolor=corlinks,
	pdfborder= 0 0 0
}

\newtheorem{theorem}{Theorem}[section]
\newtheorem{lemma}[theorem]{Lemma}
\newtheorem{claim}[theorem]{Claim}

\newtheorem{proposition}[theorem]{Proposition}

\RenewCommandCopy{\theHtheorem}{\thetheorem}

\theoremstyle{remark}
\newtheorem{preremark}[theorem]{Remark}
  {\vspace{0.5em}\begin{mdframed}[innertopmargin=-0em,skipabove=0em]\small
	\begin{preremark}\upshape}
{\end{preremark}\end{mdframed}}

\newenvironment{remarkn}[1]%
  {\vspace{0.5em}\begin{mdframed}[innertopmargin=0em,skipabove=0em]\small
	\begin{preremark}[#1]\upshape}
{\end{preremark}\end{mdframed}}

\newcommand{\bits}{\{0,1\}}

\newcommand{\bool}{\mleft\{0,1\mright\}}

\newcommand{\poly}{\mathsf{poly}}

\newcommand{\concat}{\mathrel{\circ}}
\newcommand{\eqdef}{\triangleq}

\newcommand{\ALL}{\mathsf{ALL}}
\newcommand{\corr}{\operatorname{corr}}
\newcommand{\SIZE}{\mathsf{SIZE}}
\newcommand{\nat}{\mathbb{N}}
\newcommand{\EX}{\mathsf{EX}}
\newcommand{\pmone}{\{-1,1\}}
\newcommand{\E}{\mathbb{E}}

\newcommand{\Un}{\mathcal{U}_n}

\newcommand{\Had}{\mathsf{H}}

\newcommand{\F}{\mathbb F}
\newcommand{\1}{\mathbf 1}

\definecolor{blue-violet}{rgb}{0.54, 0.17, 0.89}
\definecolor{deep-green}{rgb}{0,0.5,0}

\title{Unconditional Quantum PAC Learning Advantage in the Large Error Regime\vspace{0.6cm}}
\author{
Igor C. Oliveira\vspace{0.3cm}\thanks{E-mail:\texttt{~igor.oliveira@warwick.ac.uk}}\\{\small Department of Computer Science}\\{\small University of Warwick}\vspace{0.6cm}
}
\vspace{0.6cm}

\begin{document}

\newgeometry{left=0.5in,right=0.5in,top=1in,bottom=1in}
\maketitle

\vspace{-0.3cm}

\begin{abstract} 
We establish an unconditional super-polynomial separation between quantum and classical PAC learning of Boolean functions under the uniform distribution in the large error regime. For every advantage $\lambda(n) = n^{-\omega(1)}$, we construct a concept class that a polynomial-time quantum learner learns to agreement $1/2+\lambda(n)$ with success probability at least $1-1/n^k$. The learner uses only quantum random examples and outputs a polynomial-size deterministic classical circuit. In contrast, no polynomial-time classical learner achieves agreement $1/2 + \lambda(n)/n^k$ with success probability greater than $1/n^k$, even with membership queries.

To achieve this, we develop a nearly-optimal quantum learner that uses quantum random examples to produce a hypothesis achieving a prescribed correlation with an arbitrary Boolean function. The learner generalizes Fourier sampling from parity hypotheses to quadratic hypotheses. While standard Fourier sampling is insensitive to the sign of the correlation, we use interference to bias sampling toward positively correlated hypotheses. This is combined with a careful analysis of the resulting distribution to boost both the learner's success probability and its agreement with the target function. We believe that the resulting quantum learning algorithm might be of independent interest.

Finally, although a robust quantum PAC learning advantage holds unconditionally in the large error regime, we note that an analogous separation between quantum and classical learning in the small error regime would imply that $\mathsf{NP} \nsubseteq \mathsf{BPP}$.
\end{abstract}

\normalsize

\restoregeometry

\tableofcontents

\newpage

\section{Introduction}

We study quantum learning of Boolean functions under the uniform distribution in the \emph{large error regime}, where the required agreement with the target function is only slightly greater than $1/2$.  As explained below, even an exponentially small advantage over random guessing can be significant in this setting, both as a source of quantum learning separations and as a potential tool for proving complexity lower bounds.

\paragraph{Synergies Between Learning and Complexity.} The latter motivation comes from a general connection established by~\citep{DBLP:conf/coco/OliveiraS17, AGGOS21}. They showed that, for any typical Boolean circuit class $\mathcal C$ and choice of the  parameter $\gamma$, learning every polynomial size $\mathcal C$-circuit under the uniform distribution with agreement $1/2 + \gamma$  in quantum time
\begin{equation}\label{eq:learning_to_clbs}
        T(n) \le \frac{\gamma(n)^2\,2^n}{n^{\omega(1)}}
\end{equation}
implies $\mathsf{BQE}\nsubseteq\mathcal C[n^k]$ for every fixed $k$, where $\mathsf{BQE} \eqdef \mathsf{BQTIME}[2^{O(n)}]$, and $C[n^k]$ denotes the class of functions computable by $\mathcal{C}$-circuits of size $n^k$. This provides an approach to complexity lower bounds via quantum algorithm design.\footnote{From this perspective, quantum learning is viewed merely as a tool for separating uniform quantum computation from non-uniform Boolean circuit classes. Its usefulness for complexity theory need not depend on practical considerations of the quantum learning model, and the resulting lower bound statement makes no reference to learning algorithms.}  We note that a precedent for obtaining circuit lower bounds from algorithms is Williams' celebrated proof~\citep{Williams14} that $\mathsf{NE}\nsubseteq\mathsf{ACC}^0$ (poly-size bounded-depth circuits extended with mod $m$ gates), which uses faster classical algorithms for circuit satisfiability. In contrast, it remains open to separate $\mathsf{BQE}$ from $\mathsf{ACC}^0$, or to obtain any explicit super-polynomial circuit size lower bound against $\mathsf{TC}^0_2$ (poly-size depth-$2$ circuits extended with linear threshold gates).

In the setting of \emph{polynomial time}, the aforementioned learning connection singles out the \emph{exponentially small} correlation regime: a learning agreement of just $1/2 + n^{\omega(1)}\cdot 2^{-n/2}$ would suffice for such circuit lower bounds. This suggests the following question.
\begin{quote}
\emph{Can classical or quantum algorithms running in polynomial time learn expressive concept classes under the uniform distribution with advantage $n^{\omega(1)}\cdot 2^{-n/2}$?}
\end{quote}
Achieving even this extremely weak learning guarantee remains an intriguing open problem for prominent circuit classes such as $\mathsf{ACC}^0$ and $\mathsf{TC}^0_2$.\\

The discussion above motivates the investigation of the power of classical and quantum learners in the large error regime. Our main contribution is an unconditional separation between quantum and classical PAC learning at exponentially small advantage $\Omega(2^{-n/2})$ and, through additional ideas, a robust extension of this  separation at inverse superpolynomial advantages $n^{-\omega(1)}$. To achieve this, we develop a  quantum learning algorithm that might be of independent interest. Finally, in contrast to the large error regime, we observe that obtaining a similar quantum vs classical learning separation in the small error regime requires complexity-theoretic assumptions.

\paragraph{Quantum Learning Background.} We consider the   PAC learning model under the uniform distribution. We work with uniform quantum random examples of an unknown Boolean function $f\colon\bits^n\to\bits$, namely copies of
$$
    |\mathrm{QEX}_f\rangle
    =2^{-n/2}\sum_{x\in\bits^n}|x\rangle|f(x)\rangle.
$$
Our quantum algorithms use only these examples and make no quantum membership queries. The quantum learner outputs a deterministic classical circuit $h$.  

For Boolean functions $f,h\colon\bits^n\to\bits$, define their \emph{agreement under the uniform distribution} by
$$
    \mathsf{agreement}(f,h)
    \eqdef\Pr_{x\sim\mathcal U_n}[f(x)=h(x)].
$$
We call $\gamma$ an \emph{advantage} when the agreement is at least $1/2+\gamma$. 

It is important to distinguish two probabilities: the \emph{agreement probability} $\mathsf{agreement}(f,h)$ defined above, which captures the error of the hypothesis $h$ with respect to the target function $f$, and the \emph{success probability} of the learning algorithm, i.e., the probability that the learner produces a hypothesis with the required agreement.

For additional context and background on quantum learning theory, we refer to \citep{AW17}.\\

\begin{remarkn}{Quantum Fourier Sampling}
     Fourier sampling is a central tool for quantum learning under the uniform distribution. It allows a learner to use uniform quantum random examples to sample efficiently from the squared Fourier spectrum of the target's sign representation $F(x) \eqdef (-1)^{f(x)}$. As highlighted by~\citep{AGS21}, Fourier sampling is a common ingredient in several quantum learning algorithms. For instance, this technique and its extensions underlie quantum algorithms for learning parities~\citep{BV97}, DNF formulas under the uniform and bounded product distributions~\citep{BJ98,KRS19}, juntas~\citep{AS07}, and functions with sparse Fourier spectra~\citep{ACLPW19}. These applications make Fourier sampling a natural starting point for investigating the power of quantum examples in the large error regime.
\end{remarkn}

\vspace{0.4cm}

For arbitrary Boolean targets, Fourier sampling already gives a polynomial time learner achieving advantage $\Omega(2^{-n/2})$ with constant success probability~\citep[Appendix~A]{AGGOS21} (see \Cref{sec:fourier_sampling} for an exposition). With constant probability, the sampled parity has a Fourier coefficient of sufficiently large magnitude. However, the sampling probability depends on the square of that coefficient and therefore does not reveal its sign. Even when the magnitude is sufficiently large, randomly choosing between the parity and its negation succeeds only half the time. Thus the resulting learner's success probability is at most $1/2$. Recovering the sign of an arbitrary hypothesis with exponentially small correlation is itself costly, even with quantum membership queries (see \Cref{sec:orientationHardness}).

\paragraph{Our Approach in a Nutshell.} At this scale, the \emph{success probability} is essential to establishing a quantum learning advantage. As we explain below, a classical polynomial-time algorithm making no queries can also achieve advantage $\Omega(2^{-n/2})$ with constant success probability below $1/2$. In contrast, we show that a quantum learner can cross the success threshold $1/2$ in polynomial time, while any classical learner meeting the same guarantee requires $\Omega(2^{n/2})$ membership queries, regardless of its running time. This establishes an \emph{exponential separation} between quantum and classical learning at exponentially small advantage.

To obtain this guarantee, we develop a learner that generalizes Fourier sampling from parity hypotheses to quadratic hypotheses and uses interference to bias the sampling distribution toward positively correlated hypotheses. This allows us to show that a single quantum random example suffices to achieve the required advantage with constant success probability strictly greater than $1/2$. 

Through a careful analysis of the distribution of quadratic functions sampled by the polynomial-time quantum learner, it is possible to boost the learner's success probability and its agreement with the target function using additional random examples and larger running time. In turn, this enables us to employ a standard padding argument to translate the exponential separation between quantum and classical learners to inverse superpolynomial advantages $n^{-\omega(1)}$ in the setting of polynomial-time learning algorithms.\\

The rest of this introduction is organized as follows. We formally state our results in \Cref{sec:results} below. A detailed overview of our techniques appears in \Cref{sec:techniques}. Finally, we discuss related work in \Cref{sec:related_work}.

\subsection{Results}\label{sec:results}

In this section, we explain our results in detail. Let $\ALL_n$ denote the class of all Boolean functions
$f\colon\bits^n\to\bits$. 

\paragraph{Quantum Upper Bounds.} In our first result, we show how to modify the quantum Fourier sampling procedure in order to obtain a polynomial-time quantum learner that succeeds with probability $p^\star > 1/2$. In addition, we show that the agreement and success probabilities can be boosted using additional quantum random examples and larger running time.

\medskip

\begin{theorem}[Quantum Learning Algorithms]
\label{thm:quantum_learners}
There are universal constants
$a_0,a_1,C>0$ and $p^\star>1/2$ such that the following statements hold.

\begin{enumerate}[label=\textup{(\roman*)}]
\item
There is a quantum algorithm $\mathcal Q$ that, given
$n\in\mathbb N$ and a single quantum random example for 
$f\in\ALL_n$, runs in time $\poly(n)$ and outputs a deterministic
classical Boolean circuit $h$ of size $O(n^2)$ such that
$$
    \Pr_{\mathcal Q}\left[
        \mathsf{agreement}(f,h)
        \ge
        \frac12+a_0\cdot 2^{-n/2}
    \right]
    \ge p^\star.
$$
Concretely,\footnote{We do not optimize the constant $p^*$ in our presentation in order to keep the quantum algorithm and its analysis simpler.} one may take
$a_0=10^{-3}$ and $p^\star=0.53$.

\item
There is a quantum algorithm $\mathcal Q^+$ that, given
$n\in\mathbb N$ and $\eta\in(0,1/2)$ satisfying
$\eta\ge C\cdot 2^{-n}$, and using $O(1/\eta)$ independent quantum
random examples for $f\in\ALL_n$, runs in
time
$$
    T(n,\eta) = \poly(n)\cdot\frac1\eta
$$
and outputs a deterministic classical Boolean circuit $h$ of size at most $T(n, \eta)$ such that\footnote{The hypothesis $h$ is a depth-$3$ Boolean circuit of the form $\mathsf{MAJ}_{O(1/\eta)} \circ \oplus_{O(n^2)} \circ \mathsf{AND}_2$, where subscripts denote the fan-in of each gate type.}
$$
    \Pr_{\mathcal Q^+}\left[
        \mathsf{agreement}(f,h)
        \ge
        \frac12+
        \frac{a_1}{\sqrt{\eta}}\cdot 2^{-n/2}
    \right]
    \ge 1-\eta.
$$
\end{enumerate}
\end{theorem}

\medskip

\Cref{thm:quantum_learners} Item~(ii) gives a tradeoff between accuracy, confidence, and running time. In terms of the number $m$ of quantum random examples, we obtain advantage $\Omega(\sqrt{m/2^n})$, failure probability $O(1/m)$, and running time $m \cdot \poly(n)$. Thus an advantage $\gamma$ can be obtained in time $\poly(n)\cdot\gamma^2 \cdot  2^n$ throughout the range from a constant multiple of $2^{-n/2}$ to a sufficiently small positive constant.\footnote{For larger advantages, learning the whole truth table from $\poly(n) \cdot 2^n$ measured examples gives the same time bound up to polynomial factors.}\\

\begin{remarkn}{On the Optimality of the Parameters in \Cref{thm:quantum_learners} Item (ii)}
    Relative to the criterion of~\citep{AGGOS21} for complexity lower bounds corresponding to \Cref{eq:learning_to_clbs}, the amplified quantum learner achieves running time $\poly(n) \cdot \gamma^2 \cdot 2^n$ as opposed to $n^{-\omega(1)} \cdot \gamma^2 \cdot 2^n$. This dependence on $\gamma$ is essentially optimal for $\ALL_n$ when the learner must output a classical circuit.  Indeed, by a standard Chernoff bound, a fixed hypothesis $h$ agrees with a uniformly random Boolean function on a $1/2+\gamma$ fraction of inputs with probability at most $\exp(-2\gamma^2 2^n)$, and a simple counting argument shows that covering every target in $\ALL_n$ requires $\Omega(\gamma^2 2^n)$ bits of hypothesis description.
\end{remarkn}

\vspace{-0.2cm}

\paragraph{Classical Lower Bounds.} We show next that classical learners cannot match the guarantee in \Cref{thm:quantum_learners} Item~(i) in subexponential time. We prove an unconditional query lower bound when the target is an arbitrary Boolean function, and a conditional computational lower bound for targets computed by poly-size circuits.

For a size bound $s \colon \mathbb{N} \to \mathbb{N}$, we let $\mathsf{SIZE}[s(n)]$ denote the set of Boolean functions $f \colon \{0,1\}^n \to \{0,1\}$ computed by a Boolean circuit of size at most $s(n)$. 

The computational lower bound uses the standard cryptographic hypothesis that  one-way functions remain secure against subexponential-time classical algorithms. This assumption and its relation to strong pseudorandom functions appear, for example, in the natural proofs framework~\citep{RR97} and in the study of quantum advantages relative to efficiently computable oracles~\citep[Section~7]{AC17}. 

Formally, we say that a uniform family of functions $f_n \colon \{0,1\}^n \to \{0,1\}^{m(n)}$ computable in deterministic polynomial time is a \emph{subexponentially secure one-way function} if there is a constant $\varepsilon > 0$ such that, for every (classical) randomized algorithm $\mathcal{I}$ running in time $\leq 2^{n^{\varepsilon}}$ and for every large enough $n$, 
$$
\Pr_{x \sim \mathcal{U}_n,\;y = f_n(x),\;\mathcal{I}}[\mathcal{I}(1^n, y) \in f^{-1}_n(y)] \leq n^{-\omega(1)}.
$$

\begin{theorem}[Limitations of Classical Learners]
\label{thm:classical_limitations}
The following statements hold.

\begin{enumerate}[label=\textup{(\roman*)}]
\item
Let $\mathcal A$ be a randomized algorithm that makes at
most $q$ membership queries to $f\in\ALL_n$ 
and outputs a deterministic classical Boolean circuit $h$.\footnote{This
statement places no restriction on the running time of $\mathcal A$
or on the size of $h$.} For every $\gamma>q/2^{n+1}$, there is a function  
$f\in\ALL_n$ such that
$$
    \Pr_{\mathcal A}\left[
        \mathsf{agreement}(f,h)
        \ge
        \frac12+\gamma
    \right]
    \le
    \frac12.
$$
In particular, when learning $\ALL_n$ it is not possible to achieve correlation $\Omega(2^{-n/2})$ with success probability $> 1/2$ using $o(2^{n/2})$ classical membership queries.
\item
Assume the existence of subexponentially secure one-way functions. There is $k\ge1$ such that, for every pair of
 constants $a_0>0$ and $p^\star>1/2$,  no randomized algorithm $\mathcal A$ has the
following property:

\medskip
\noindent For every large enough $n\in\mathbb N$ and every 
 $f\in\mathsf{SIZE}[n^k]$, 
the algorithm runs in time $o(2^{n/2})$, makes classical membership
queries, and outputs a deterministic classical Boolean circuit $h$  satisfying
$$
    \Pr_{\mathcal A}\left[
        \mathsf{agreement}(f,h)
        \ge
        \frac12+a_0\cdot 2^{-n/2}
    \right]
    \ge p^\star.
$$
\end{enumerate}
\end{theorem}

In other words, under the existence of subexponentially secure OWFs, it is not possible to achieve correlation $\Omega(2^{-n/2})$ and constant success probability $p^\star > 1/2$ for every $f \in \mathsf{SIZE}[n^{O(1)}]$  in time $o(2^{n/2})$.

\paragraph{Quantum PAC Learning Advantage.} Together, the two theorems give an \emph{unconditional} exponential separation between one quantum random example and $\Omega(2^{n/2})$ classical membership queries, and a conditional separation between quantum polynomial time and classical time $o(2^{n/2})$ for a fixed polynomial-size circuit class. Note that the computational separation requires no specific number-theoretic assumption, such as the hardness of factoring or computing discrete logarithms. In particular, our results show that an exponential quantum vs classical PAC learning separation for a concept class $\mathfrak{C} \subseteq \mathsf{SIZE}[\poly]$ can be obtained from a \emph{generic cryptographic assumption}.

Next, we build on the results described above to translate the exponential quantum advantage at the exponentially small correlation regime into a robust polynomial advantage in the agreement regime $1/2 + \lambda(n)$, for any ``constructive'' function $\lambda(n) = 1/n^{\omega(1)}$. We say that a function $\lambda \colon \mathbb{N} \to \mathbb{R}^+$ is constructive if $\lambda(n)$ can be  computable from $1^n$ in time $\poly(n)$.\footnote{It is enough that there is an efficient algorithm that approximates $\lambda(n)$. We refer to \Cref{sec:padded-separation} for the details.} For example, \Cref{cor:padded-separation} below applies to $\lambda(n)=n^{-\log n}$, $\lambda(n)=n^{-\log\log n}$, $\lambda(n)=n^{-\log^* n}$, etc.

\begin{theorem}[Robust Quantum PAC Learning Advantage at Inverse Superpolynomial Correlation]\label{cor:padded-separation}
Let $\lambda(n)$ be any constructive function satisfying 
$
n^{\omega(1)} \cdot 2^{-n/2} \leq \lambda(n) \leq n^{-\omega(1)}
$, and let $k \geq 1$ be an arbitrary constant.
Then there is a concept class $\mathfrak C=\{\mathcal C_n\}_{n\ge1}$ for which the following holds.
\begin{itemize}
    \item  
There is a quantum algorithm $\mathcal A$ that, for every sufficiently large $n$ and every $f\in\mathcal C_n$, uses quantum random examples, runs in $\poly(n)$ time, and outputs a deterministic classical circuit $h$ satisfying
$$
    \Pr_{\mathcal{A}}\left[
        \mathsf{agreement}(f,h)
        \ge\frac12+\lambda(n)
    \right]\ge1-\frac{1}{n^k}.
$$
\item For every randomized membership-query polynomial-time learner $\mathcal B$, for every sufficiently large $n$ there is $f\in\mathcal C_n$ such that
$$
    \Pr_{\mathcal B}\left[
        \mathsf{agreement}(f,h)
        \ge\frac12+\frac{\lambda(n)}{n^k}
    \right]\le\frac{1}{n^k}.
$$
\end{itemize}
Moreover, if subexponentially secure one-way functions exist, the concept class $\mathfrak C$ can be chosen to satisfy $\mathfrak C\subseteq\mathsf{SIZE}[\poly]$.
\end{theorem}

To our knowledge, \Cref{cor:padded-separation} establishes the first unconditional super-polynomial separation in the time complexity of PAC learning Boolean functions under the uniform distribution. The quantum learner runs in polynomial time, uses only quantum random examples, and outputs a classical hypothesis, while no polynomial-time classical learner achieves the same guarantee even with membership queries. Moreover, the separation is robust, in the sense that a polynomial-time classical learner is not able to achieve a weaker agreement guarantee under a significantly lower success probability.\\

\begin{remarkn}{The Small Error Regime}
    We remark that obtaining a similar robust separation in the agreement regime $1/2+\lambda(n)$, where $\lambda(n)\ge 1/\poly(n)$, with classical circuit hypotheses would imply that $\mathsf{NP}\nsubseteq\mathsf{BPP}$. In this sense, \Cref{cor:padded-separation} is essentially optimal for unconditional separations of this form, given our current knowledge of complexity theory.

The reason is that polynomial-time agnostic learning of polynomial-size circuits is possible under the average-case easiness assumption $\mathsf{DistNP}\subseteq\mathsf{AvgBPP}$ \citep{DBLP:conf/coco/GoldbergKLO22}. We use two features of our separation: the quantum learner outputs a deterministic classical circuit, and classical hardness persists at the lower agreement $1/2+\lambda(n)/2$.

In more detail, let $\mathfrak{C}=\{\mathcal{C}_n\}_{n\ge 1}$ be an arbitrary class of Boolean functions. Suppose that $\mathfrak{C}$ admits a polynomial-time quantum learner that, for every sufficiently large $n$ and every $f\in\mathcal{C}_n$, outputs a deterministic classical circuit $h$ with $\mathsf{agreement}(f,h)\ge 1/2+\lambda(n)$ with probability at least $2/3$, where $\lambda(n)\ge 1/n^a$ for some fixed $a>0$. The running time bound implies that, for some fixed $\ell>0$, every such $f$ admits a circuit $D_f$ of size at most $n^\ell$ achieving this agreement. Under $\mathsf{DistNP}\subseteq\mathsf{AvgBPP}$, \citep[Theorem 56]{DBLP:conf/coco/GoldbergKLO22} gives a classical agnostic learner for $\mathsf{SIZE}[n^\ell]$ under the uniform distribution, running in time $\poly(n,1/\varepsilon)$. Using membership queries to obtain uniform labeled examples, and choosing $\varepsilon=1/(2n^a)$, this learner outputs, with probability at least $2/3$, a hypothesis whose agreement with $f$ is at least the maximum agreement achievable by a circuit in $\mathsf{SIZE}[n^\ell]$ minus $\varepsilon$, and therefore at least $1/2+\lambda(n)/2$. This contradicts the robust classical lower bound. Thus such a separation implies $\mathsf{DistNP}\nsubseteq\mathsf{AvgBPP}$, and hence $\mathsf{NP}\nsubseteq\mathsf{BPP}$.
\end{remarkn}

\medskip

\subsection{Techniques}\label{sec:techniques}

We now describe our main ideas and techniques in more detail, including how they can be combined to derive \Cref{cor:padded-separation}. First, we review quantum Fourier sampling and its limitations, then explain how to achieve a similar guarantee using a classical algorithm. While neither approach achieves  success probability higher than $1/2$, these considerations reveal useful insights that are explored in our results.

\paragraph{Review of Quantum Fourier Sampling and Its Limitations.} As mentioned above, a standard technique in quantum learning theory is \emph{Fourier sampling}. Let $f \colon \{0,1\}^n \to \{0,1\}$ be an unknown function, and suppose we have access to a quantum random example
$$
|\mathrm{QEX}_f\rangle
=
\frac{1}{\sqrt N}\sum_{x\in\bits^n}|x\rangle|f(x)\rangle,
$$
where $N \eqdef 2^n$.
Let $F(x)\eqdef(-1)^{f(x)}$ be the sign representation of $f$. Then, an efficient  computation allows us to produce the state 
$$
\sum_{S\in\bits^n}
\widehat F(S)|S\rangle,
$$
where $\widehat F(S) \in [-1,1]$ is the \emph{correlation} between $F$ and the parity function $\chi_S$. This means that a measurement in the computational basis  samples a parity $\chi_S(x)=(-1)^{S\cdot x}$ with probability $\widehat F(S)^2$. Using that the $N = 2^n$ squared Fourier coefficients sum to $1$, it is not hard to show that the sampled parity has correlation of magnitude $\Omega(N^{-1/2})$ with constant probability. We refer to \Cref{sec:fourier_sampling} for the details.

The difficulty is its \emph{orientation}: the parity may be positively or negatively correlated with the target. Squaring the Fourier coefficient erases its sign, and the phase states for $F$ and $-F$ differ only by a global phase. Consequently, the sampling outcome cannot distinguish these two targets. Choosing uniformly between the sampled parity and its negation gives the required positive correlation with constant success probability, but this probability is always at most $1/2$. Moreover, at exponentially small correlation, recovering the correct orientation by a separate test is expensive even with quantum membership queries (see \Cref{sec:orientationHardness}).

\paragraph{An Efficient Classical Learner with Success Probability at Most $\bm{1/2}$.}
The same correlation scale can be reached classically \emph{without any queries}: simply output a uniformly random truth table. For any fixed target, its correlation is the average of $N$ independent unbiased signs. The sum fluctuates on the scale $\sqrt N$, so positive correlation $\Omega(N^{-1/2})$ occurs with constant probability. However, symmetry again bounds this success probability by $1/2$. Moreover, a random truth table requires $N=2^n$ bits to describe, so this observation alone does not give an \emph{efficient learner} with a succinct hypothesis.

To obtain a polynomial-time learner, we (partially) derandomize the truth table construction using limited independence, a standard technique from the theory of computational pseudorandomness (see, e.g., \citep{DBLP:journals/fttcs/Vadhan12}). We say that a family $\mathcal{H} = \{h_a\}_{a \in \{0,1\}^r}$ of functions $h_a \colon \{0,1\}^m \to \{0,1\}$ is $k$\emph{-wise independent} if, for a uniformly random seed $a \sim \{0,1\}^r$, its values at any set of at most $k$ distinct inputs are independent uniform bits. While the second moment identifies the natural scale of correlation, it could be carried by a few exceptionally large values while almost all hypotheses have correlation zero. In contrast, one can show that a fourth moment bound  rules out this scenario, ensuring that useful magnitudes occur with constant probability. The important point is that full independence is unnecessary for these calculations: a $4$-wise independent family $\mathcal{H}$ suffices to reproduce both moments. 

In more detail, for $H_a=(-1)^{h_a}$ and a fixed target $F$, it is possible to show using the $4$-wise independence of $\mathcal{H}$ that the normalized correlation $Z_a$ satisfies
$$
\E_a [Z_a^2]=1 \quad \text{and}
    \quad \E_a [Z_a^4]\leq 3,\quad \text{where} \quad 
    Z_a\eqdef\sqrt N\,\corr(H_a,F)
        =\frac1{\sqrt N}\sum_xF(x)H_a(x).
$$
Then, applying the Paley-Zygmund Inequality (\Cref{lem:PZ-Inequality}) to $Z_a^2$ gives a constant lower bound on $\Pr_a[|Z_a|\ge1/2]$. If the family also contains each hypothesis and its negation with equal weight, half of this probability comes from \emph{positive} correlations. 

An explicit $4$-wise independent family with polynomial length seeds and polynomial size circuits therefore gives an efficient classical learner making zero queries, whose success probability is bounded below by a positive constant and above by $1/2$ (\Cref{prop:zero-query-construction}). While the learner still chooses a polynomial size random seed, this partial derandomization reduces the randomness and hypothesis size, as desired.\\

Before we describe how to go beyond success probability $1/2$ for advantage $\gamma = \Omega(2^{-n/2})$ using a quantum learner, we briefly explain why a classical randomized learner that makes subexponentially many membership queries cannot attain this guarantee.

\paragraph{Why $\bm{o(2^{n/2})}$ Classical Queries Cannot Cross the $\bm{1/2}$ Barrier.} The proof of \Cref{thm:classical_limitations} Item~(i) follows a standard query lower bound argument. As above, we use $F$ and $H$ to denote the sign representations of Boolean functions $f$ and $h$, respectively.  Choose the target $f$ uniformly from $\ALL_n$, fix the learner's random coins, and condition on its complete query transcript. Even when the queries are adaptive, the labels at unqueried inputs remain independent uniform bits, and the output hypothesis $h$ is fixed. The queried inputs contribute at most $q$ to the sum defining $N \cdot \corr(H,F)$, and the remaining contribution is a sum of independent unbiased signs, hence symmetric about zero.

Achieving agreement $1/2+\gamma$ requires this total sum to be at least $2N\gamma$. If $2N\gamma>q$, success therefore requires the unseen contribution to be strictly positive, an event of probability at most $1/2$. Averaging over transcripts, coins, and targets yields a target on which the learner's success probability is at most $1/2$. In particular, $q=o(\sqrt N)$ cannot support a guarantee above $1/2$ at advantage $\Omega(N^{-1/2})$. Intuitively, the observed labels can shift the correlation by only $q/N$, which is too small to overcome the uncertainty on the relevant scale.

\paragraph{Correlation-Biased Quantum Sampling and Quadratic Hypotheses.}
To prove \Cref{thm:quantum_learners} Item~(i), we combine the succinct  hypotheses of the classical learner with an extension of the ideas behind Fourier sampling. 

The key is to retain the known branch that the usual Fourier procedure discards when preparing a phase state. To make this explicit, write
$$
    |u\rangle\eqdef\frac1{\sqrt N}\sum_x|x\rangle,
    \qquad
    |F\rangle\eqdef\frac1{\sqrt N}\sum_xF(x)|x\rangle.
$$
Applying a Hadamard gate to the label qubit of a quantum example gives
$$
    |\mathrm{QEX}_f\rangle
    \longmapsto
    \frac1{\sqrt2}\bigl(|u\rangle|0\rangle+|F\rangle|1\rangle\bigr).
$$
The branch labelled $0$ is the known uniform state, independent of the target, while the branch labelled $1$ contains the target values as phases. The usual Fourier procedure measures the label qubit and keeps the outcome $1$, which occurs with probability $1/2$. It then uses the isolated phase state $|F\rangle$. Since $|{-F}\rangle=-|F\rangle$, replacing the target by its complement changes this state only by a global phase, which cannot affect any subsequent measurement statistics.

In the joint state, however, replacing $F$ by $-F$ changes only the second summand: the reference branch $|u\rangle|0\rangle$ stays fixed. The sign change is therefore a \emph{relative phase} between the two branches, rather than a global phase of the entire state. This preserves the distinction between the two orientations, but retaining the branches is only the first step. To make that distinction affect the output distribution, we must keep the label qubit coherent and later transform both branches into a common measurement basis, so that their amplitudes interfere. Measuring the label in the computational basis at this stage, or discarding it, would remove the coherence between the branches and lose this information.

The known branch is especially convenient as a coherent reference because it can be transformed into a single fixed basis state. Place the label qubit first and apply a Hadamard transform to the input register only when the label is $0$. This sends $|0\rangle|u\rangle$ to $|0\rangle|0^n\rangle$, while leaving the target branch unchanged. Calling this fixed reference state $|\mathord{\perp}\rangle$ and writing $|x\rangle_{\mathrm{lab}}=|1\rangle|x\rangle$, we obtain
$$
    \frac1{\sqrt2}|\mathord{\perp}\rangle
    +\frac1{\sqrt{2N}}\sum_xF(x)|x\rangle_{\mathrm{lab}}.
$$
We note that this preparation uses the original quantum example without any measurement or additional access to the target Boolean function. 

Jumping ahead, a suitable quantum transform $W$ described below maps the entire label space into a hypothesis register, with no separate branch label retained in an unobserved register. The reference and target contributions to each hypothesis amplitude can then add coherently, allowing their relative sign to favor the positively correlated orientation.

In more detail, our hypothesis family $\mathcal{H} = \{h_a\}_{a \in \{0,1\}^r}$ consists of all quadratic Boolean polynomials
$$
    h_a(x)=a_{\emptyset}
        \oplus\bigoplus_i a_ix_i
        \oplus\bigoplus_{i<j}a_{ij}x_ix_j.
$$
A seed contains $r=1+n+\binom n2$ coefficients, and each hypothesis has a circuit of size $O(n^2)$. While affine functions provide pairwise independence, it is possible to show that the quadratic function family $\mathcal{H}$ is $4$-wise independent. The details appear in \Cref{sec:four-wise-family}.

Write $M=2^r$ for the number of seeds. The aforementioned quantum transform $W$ converts the reference branch into a uniform superposition of seeds and the target branch into a superposition whose amplitude at seed $a$ is proportional to the normalized correlation $Z_a$ defined above. Up to a common normalization, these contributions are $1$ and $Z_a$. Keeping the two branches coherent adds their amplitudes before measurement, giving amplitude $(1+Z_a)/\sqrt{2M}$ and hence the exact output distribution
$$
    \mathsf P_F(a)=\frac{(1+Z_a)^2}{2M}.
$$

We note that the reference branch alone would sample hypotheses uniformly, just as the classical learner does. The target branch alone would assign weights proportional to $Z_a^2$, generalizing Fourier sampling to quadratics but still erasing the sign. Together they produce $(1+Z_a)^2=1+2Z_a+Z_a^2$. Crucially, the linear  term $2Z_a$ favors positive correlations: for opposite hypotheses with normalized correlations $z$ and $-z$, where $z>0$, the weights differ by $(1+z)^2-(1-z)^2=4z$. The correct orientation is therefore more likely to be sampled.

To complete the proof sketch of \Cref{thm:quantum_learners} Item (i), we briefly discuss the implementation of $W$ and the achieved success probability. The quadratic family $\mathcal{H}$ described above supports this transform efficiently using a Hadamard transform over its coefficient register. We refer to the full construction  in \Cref{sec:quantum_learner}. The remaining issue is to show that the preference for positive correlations is strong enough to raise success above $1/2$. Here the fourth moment bound $\E_a [Z_a^4]\leq 3$ stated above has its second role: it controls the second moment of the positive part $X \eqdef\max\{Z_a,0\}$ under the reweighted distribution $\mathsf P_F$. Together with symmetry and $\E_a[Z_a^2]=1$, it gives $\E_{\mathsf P_F}[X]\ge1$ and $\E_{\mathsf P_F}[X^2]\le1+\sqrt3/2<2$. The Paley-Zygmund Inequality then yields $\Pr_{\mathsf P_F}[Z_a\ge c]>1/2$ for a sufficiently small fixed $c>0$, as desired. Thus the same moment bounds both supply useful classical candidates and turn interference into a quantum success guarantee above $1/2$.\\

\begin{remarkn}{Limits of Parity Functions and of Pairwise Independence}
Quadratic functions play an important role in the analysis of our learning algorithms. For the interested reader, in \Cref{sec:whyParitiesDoNotSuffice} we provide a concrete example to illustrate why the use of linear functions and pairwise independence is insufficient to achieve the guarantees on the success probabilities of the classical and quantum learners described above. 
\end{remarkn}

\paragraph{Amplifying Correlation and Success Probability.}
For \Cref{thm:quantum_learners} Item~(ii), we run the quantum learner independently on several examples and output the majority of its hypotheses. Note that a success probability above $1/2$ for the overall correlation does not by itself justify this step: hypotheses with positive correlation may still make systematic errors on the same inputs. Nevertheless, the quantum sampler provides the stronger property needed here. At \emph{every input}, a sampled hypothesis has the same small positive bias toward the correct value, and its agreement signs at distinct inputs are independent in pairs. Repetition increases each pointwise bias, while independence across input pairs makes the overall correlation concentrate around its positive mean.

Specifically, under the sampler's output distribution $\mathsf P_F(a)$, which corresponds to an output hypothesis $h_a$, the signs $Y_x\eqdef F(x)H_a(x)$ satisfy $\E_{\mathsf P_F}[Y_x]=N^{-1/2}$, and $Y_x,Y_y$ are independent whenever $x\ne y$. These properties follow from $4$-wise independence of the original family $\mathcal{H}$ (see \Cref{lem:pointwise-bias}). Now take an odd number $m\le N$ of independent runs. At a fixed input, majority turns $m$ independent signs of mean $N^{-1/2}$ into a sign of mean $\Omega(\sqrt{m/N})$ (see \Cref{lem:majority-bias}). Independence across pairs of inputs can be shown to pass to their majority outputs, so the overall correlation has variance at most $1/N$. Its mean grows with $m$, while its variance stays bounded by the same quantity. Chebyshev's inequality consequently gives correlation $\Omega(\sqrt{m/N})$ with failure probability $O(1/m)$. Finally, choosing $m=\Theta(1/\eta)$ yields \Cref{thm:quantum_learners} Item~(ii), with time and circuit size $\eta^{-1}\poly(n)$.

\paragraph{From Exponentially Small to Inverse Superpolynomial Advantage.}
We obtain \Cref{cor:padded-separation} by a standard padding argument and an extension of the classical lower bound. Set $r_n=2^{k\lceil\log_2 n\rceil}$, so that $n^k\le r_n=n^{O(1)}$. For the desired advantage $\lambda(n)$, the concept class consists of target functions $f_g(x,z)=g(x)$ that depend on only $d$ of their $n$ input bits, with $2^d=\Theta(r_n^3\lambda(n)^{-2})$. The assumptions on $\lambda$ ensure that $d=\omega(\log n)$ and $d<n$ eventually, while constructivity allows $d$ to be computed efficiently.

For the quantum upper bound, it is not hard to see that quantum random examples for $g$ can be obtained from quantum random examples for $f_g$. We apply the amplified quantum learner (\Cref{thm:quantum_learners} Item~(ii)) with $\eta=1/(Lr_n^3)$, for a sufficiently large fixed constant $L$. Using $O(r_n^3)$ examples in polynomial time gives advantage $a_1\sqrt{Lr_n^3/2^d}\ge\lambda(n)$, with success probability at least $1-\eta\ge1-1/n^k$.

Classically, the relevant transcript lower bound argument now concerns the $2^d$ independent ``core'' labels corresponding to a random function $g \colon \{0,1\}^d \to \{0,1\}$. A polynomial number of queries is negligible compared with $2^d\lambda(n)/r_n=\Theta(r_n^2/\lambda(n))=n^{\omega(1)}$, and one can show that some target has success probability at most $1/2$ even at advantage $\lambda(n)/r_n\le\lambda(n)/n^k$. This remains true even if the learner's hypothesis depends on all padding bits: averaging over those bits assigns each core input a weight in $[-1,1]$, preserving symmetry of the unseen contribution.

To obtain a vanishing success probability, we also control the size of the unseen contribution. The unknown core labels are independent and unbiased, so their contributions largely cancel, leaving fluctuations on a scale at most $2^{d/2}$. A simple calculation shows that our choice of parameters makes success at advantage $\lambda(n)/r_n$ require a favorable fluctuation larger than this scale by a factor of order $\sqrt{r_n}$. The polynomially many queried labels contribute too little to close this gap. Formally, using Chebyshev's inequality and our choice of constants it can be shown that the average success probability is at most $1/(16r_n)\le1/n^k$. Averaging therefore supplies a target with this bound, which also holds at the required advantage $\lambda(n)/n^k\ge\lambda(n)/r_n$.\\

Finally, it is possible to obtain the same quantum learning advantage for a concept class whose functions admit polynomial-size Boolean circuits, under a generic cryptographic assumption.

\paragraph{Transferring the Separation to Polynomial Size Concepts.}
It is well known that one-way functions (OWFs) yield pseudorandom function families (PRFs) with polynomial-size circuits~\citep{HILL99,GGM86}. Intuitively, a classical learner running within the security bound cannot distinguish a random function from a pseudorandom one, so its limitations on random targets should also apply to targets from the pseudorandom family. The quantum upper bounds already apply to every Boolean function, including these efficiently represented functions. The task is therefore to use the PRF family to transfer the classical lower bounds to a concept class contained in $\SIZE[\poly]$.

The subtlety here is \emph{verification}.  To turn a successful learner into a distinguisher, we must check whether its output has the promised positive correlation. An exponentially small correlation cannot in general be certified in polynomial time, so ordinary polynomial security does not justify this reduction.  To address this issue, we use subexponential security and a change of scale dating back to the natural proofs argument~\citep{RR97}, i.e., we choose the target's relevant input dimension sufficiently small relative to the cryptographic security parameter that exhaustive correlation testing fits within the security bound, while the resulting functions still have polynomial size circuits. A classical randomized learner crossing the success probability threshold would then give a distinguisher with suitable advantage. This idea and its extensions underlie    \Cref{thm:classical_limitations} Item~(ii) and the conditional part of \Cref{cor:padded-separation}.

\subsection{Related Work}\label{sec:related_work}

Servedio and Gortler~\citep{SG04} established polynomial relations between classical and quantum sample complexity for distribution-free PAC learning, and between their query complexities for exact learning with membership queries. For exact learning on $n$ input bits, $Q$ quantum queries can be simulated using $O(nQ^3)$ deterministic classical queries. Recent constructions match this bound and its sharper randomized analogue~\citep{AGM26}. For distribution-free PAC learning in the usual small error regime, Arunachalam and de Wolf~\citep{AW18} proved that classical and quantum sample complexities coincide up to constant factors.

Under a fixed example distribution, advantages are possible. Uniform parity learning uses $O(1)$ quantum examples versus $\Omega(n)$ classical examples at fixed small error and constant confidence~(see, e.g.,~\citep[Section 5.3]{AW17}). With stronger oracle access, namely a quantum example preparation unitary and its inverse, a nearly quadratic improvement in the accuracy dependence is possible even for distribution-free learning~\citep{SSG24}. In contrast, our exponential sample complexity gap concerns the large error regime: under the uniform distribution, one quantum random example yields advantage $\Omega(2^{-n/2})$ with success probability above $1/2$, whereas exponentially many classical membership queries are necessary. 

In the computational setting, Servedio and Gortler~\citep{SG04} proved a PAC learning separation assuming classical hardness of factoring Blum integers, i.e., they describe a concept class that is efficiently quantum learnable under arbitrary distributions but is not efficiently classically PAC learnable. Their quantum learner uses Shor's factoring algorithm and ordinary classical examples, and outputs a classical circuit. Liu, Arunachalam, and Temme~\citep{LAT21} give a quantum kernel method that achieves $99\%$ accuracy with success probability at least $2/3$ for a classification problem under the uniform distribution on $\mathbb{Z}_p^\ast$. Assuming the hardness of discrete logarithm, no classical polynomial-time learner achieves inverse polynomial advantage over random guessing with the same confidence. Jerbi et al.~\citep{JGMMD24} also obtain a computational PAC learning advantage under the uniform distribution assuming the discrete cube root assumption, with quantum computation used during training and entirely classical evaluation of the resulting hypothesis.

These computational separations exploit arithmetic structure. Under the generic assumption that one-way functions exist, Servedio and Gortler~\citep{SG04} established a separation between quantum and classical exact learning from membership queries, where successful learning requires a hypothesis agreeing with the target on every input. This result does not establish a PAC learning separation from random examples.

Going beyond the standard PAC and exact learning frameworks, Gavinsky~\citep{Gav12} obtained an unconditional exponential separation between quantum and classical learning in a predictive model for relations, where an input may have several valid outputs. Their learner retains a quantum state to answer prediction queries, while our results concerns Boolean functions and outputs a classical hypothesis. Conditional quantum learning advantages have also been established for noisy parity and linear equations~\citep{CSS15,GKZ19}, and for generative and density learning of classical distributions under cryptographic assumptions~\citep{SSHE21,PSES23}. For learning certain properties of quantum states and processes, unconditional exponential separations in sample or query complexity arise between learners with quantum memory and those retaining only classical information between samples or experiments~\citep{HKP21,CCHL21}.

Our quantum sampler builds on Fourier sampling~\citep{BV97,BJ98}, retaining  the known branch of a labeled quantum example as a coherent reference, so that measurement probabilities contain a term linear in signed correlation. Limited independence supplies a succinct quadratic family with controlled moments, giving positive correlation with success probability strictly above $1/2$. A close precedent is van Dam's approximate oracle interrogation~\citep{vD98}, where one quantum query produces a string of $N$ bits with expected agreement $1/2+1/(2\sqrt N)$ with the oracle string. This construction already biases the output toward positive signed correlation. The main distinction is computational: van Dam's algorithm outputs a full truth table of length $N=2^n$, whereas our sampler produces a succinct classical hypothesis in time polynomial in $n$. The use of a coherent reference has conceptual precedents in phase retrieval, where a known reference makes intensity measurements sensitive to relative phase through interference terms~\citep{KH90,BSLLC19}. A related use of interference appears in quantum classification, where measurement probabilities depend on inner products between encoded data vectors~\citep{SFP17}.

The use of quadratic hypotheses connects our construction to quadratic Goldreich-Levin algorithms, initiated by Tulsiani and Wolf~\citep{TW14}. Using classical membership queries, recent algorithms find quadratic phases with absolute correlation at least $\tau-\varepsilon$ in time polynomial in $n$ and quasi-polynomial in $1/\varepsilon$~\citep[Theorem~1.5]{BCADG26}, or at least $\tau^C$ in time polynomial in $n$ and $1/\tau$, for a universal constant $C>1$~\citep[Theorem~1.3]{ADGG26}, where $\tau$ is the maximum absolute quadratic correlation. Even with classical membership queries, these bounds do not yield a polynomial-time learner at the exponentially small parameters considered here.

\vspace{0.35cm}

\noindent\rule{\textwidth}{0.4pt}

\vspace{0.05cm}

\paragraph{Acknowledgements.} We thank Rahul Ilango for reminding us of the simple but insightful observation that outputting a random function gives a classical learner with a guarantee similar to the quantum learner based on Fourier sampling \citep{EmailThread}. Appendix A of \citep{AGGOS21} and this observation served as the starting point for our investigation. We also thank Tom Gur for an email exchange on previous work related to quantum learning advantage.

\paragraph{Use of AI Tools.} Since \citep{AGGOS21} and \citep{EmailThread}, we had believed, perhaps naively, that the uncertainty in the sign of the hypothesis produced by Fourier sampling was an inherent limitation of any efficient quantum learner, just as it is for randomized learners. Generative AI led us to recognize that this intuition was incorrect. The central idea behind the correlation-biased quantum sampler, which combines a derandomized classical learner based on anti-concentration with a variant of quantum Fourier sampling, was discovered by GPT-5.5. We also used GPT-5.5, GPT-5.6, GPT-6, and GPT-6.1 extensively to explore several related directions, prepare initial drafts of the proofs of auxiliary results, and for general assistance with writing and verification. These tools materially affected all sections of the paper. We take full responsibility for the correctness of the paper and exposition of the results.

\paragraph{Funding Information.} This work received support from the UKRI Frontier Research Guarantee Grant EP/Y007999/1 and the Centre for Discrete Mathematics and its Applications (DIMAP) at the University of Warwick. The author would also like to thank the Isaac Newton Institute for Mathematical Sciences, Cambridge, for support and hospitality during the programme Logical Foundations of Complexity Theory (LFC), where part of the work on this paper was undertaken. The LFC programme was supported by EPSRC grant EP/Z000580/1.

\section{Preliminaries}

\subsection{Basic Notation}\label{sec:basic_notation}

For a given $n \geq 1$ we will implicitly use $N = 2^n$, unless stated otherwise. 

\paragraph{Probabilities.} We use $x \sim S$ to denote that $x$ is sampled uniformly from the set $S$. We often use $\mathcal{U}_n$ to denote the uniform distribution over $\{0,1\}^n$.  We will sometimes use boldface letters to emphasise a random choice, e.g., to distinguish each function $h_a$ in a family indexed by $a \in A$ and a random function $h_{\bm{a}}$, where $\bm{a} \sim A$. For an event $E$, we use $\bm{1}_E$ to denote its indicator random variable. We use $E^{\mathsf c}$ to denote the complement of an event $E$.

\paragraph{Function Classes.} Let $\ALL_n$ denote the class of all  functions
$f\colon\bits^n\to\bits$. For a size bound $s \colon \mathbb{N} \to \mathbb{N}$, we let $\mathsf{SIZE}[s(n)]$ denote the set of Boolean functions  computed by a Boolean circuit of size at most $s(n)$.

\paragraph{Representation of Boolean Functions.} We will often use lowercase letters for Boolean-valued functions and the
corresponding uppercase letters for their sign representations.  Thus a target
\[
    f\colon\bits^n\to\bits
\]
has \emph{sign representation}
\[
    F\colon\bits^n\to\pmone,\qquad F(x)=(-1)^{f(x)}.
\]
Similarly, a Boolean-valued hypothesis $h\colon\bits^n\to\bits$ has sign representation $H(x)=(-1)^{h(x)}.$

\paragraph{Agreement and Correlation.} We let
\[
    \corr(H,F) \eqdef \E_{x\sim\Un}[H(x)F(x)] \in [-1,1]
\]
 denote the \emph{correlation} between $H$ and $F$. Agreement and correlation are related by
\begin{equation}\label{eq:agreement_vs_correlation}
    \Pr_{x\sim \Un}[h(x)=f(x)]
    =\frac12+\frac12\corr(H,F).
\end{equation}

\subsection{Classical Learning Theory}

In this section, we review the standard PAC learning model, fix notation, and make our conventions explicit. We refer to a textbook such as \citep{KearnsVazirani:94} for additional information.

In the PAC learning framework, the learning target is an unknown Boolean function $f \colon \bool^n \to \bool$, called a \emph{concept}. A \emph{concept class} $\mathcal{C}$ is a family of such functions. For each $n \in \nat$, we write
$$
\mathcal{C}_n \eqdef \mathcal{C} \cap \{f \colon \bool^n \to \bool\}
$$
for the restriction of $\mathcal{C}$ to inputs of length $n$.

Learnability is defined relative to a class $\mathcal{D}$ of distribution ensembles. Each ensemble $D = \{D_n\}_{n \in \nat}$ consists of probability distributions $D_n$ supported on $\bool^n$. Given a target concept $f \in \mathcal{C}_n$ and a distribution $D_n$, a \emph{labelled example oracle} $\EX(f,D_n)$ returns, on each invocation, an independent pair $(x,f(x))$, where $x$ is sampled according to $D_n$.

\paragraph{PAC Learning Model. }
Let $\mathcal{C}$ be a concept class and let $\mathcal{D}$ be a class of distribution ensembles. An algorithm $A$ \emph{PAC learns} $\mathcal{C}$ over $\mathcal{D}$ if, for every $D \in \mathcal{D}$, every $\varepsilon,\delta \in (0,1)$, every sufficiently large $n \in \nat$, and every $f \in \mathcal{C}_n$,
$$
\Pr_{A,\,\EX(f,D_n)}
\left[
\Pr_{x \sim D_n}\bigl[h(x)=f(x)\bigr] \geq 1-\varepsilon \;\; \text{for} \;\; 
  h \leftarrow A^{\EX(f,D_n)}
\bigl(1^n,1^{\lceil \varepsilon^{-1}\rceil},1^{\lceil \delta^{-1}\rceil}\bigr)   \right]
\geq 1-\delta.
$$
The outer probability is taken over the internal randomness of $A$ and the independent samples returned by $\EX(f,D_n)$. In other words, except with probability at most $\delta$, the hypothesis produced by the learner has error at most $\varepsilon$ under $D_n$.\\

We will only consider learning algorithms over the \emph{uniform distribution}, i.e., we assume $D_n = \mathcal{U}_n$ throughout the paper.

A PAC learner is \emph{proper} if it always outputs a hypothesis $h \in \mathcal{C}_n$. A learner whose output need not belong to $\mathcal{C}_n$ is called \emph{improper}. All learning algorithms considered in this paper are improper.

In some results considered in this paper, the \emph{confidence} parameter $\delta$ and the \emph{accuracy} parameter $\varepsilon$ are fixed. In such cases, the parameters are not provided as inputs to the learner. 

\paragraph{Membership Queries.} In the PAC model with \emph{membership queries}, the learner additionally receives oracle access to the target concept $f$. It may therefore choose any $x \in \bool^n$ and query the oracle for the value $f(x)$. Note that a randomized learning algorithm operating under the uniform distribution can simulate each example oracle call $\EX(f,\mathcal{U}_n)$ using a membership query. For this reason, when proving lower bounds on classical learning algorithms we will  restrict our attention to membership queries.

\subsection{Quantum Learning Algorithms}

We consider quantum learning algorithms with access to \emph{quantum random examples} under the uniform distribution. We describe the model below, and refer to \citep{AW17, AGS21} for more information related to quantum learning theory and quantum random examples.

\paragraph{Learning from Quantum Random Examples.} Under the uniform distribution, a classical random example is a pair
$(x,f(x))$, where $x$ is sampled uniformly from $\bits^n$. Its quantum
analogue is the coherent state
$$
    |\mathrm{QEX}_f\rangle
    =
    \frac{1}{\sqrt{2^n}}
    \sum_{x\in\bits^n}|x\rangle|f(x)\rangle.
$$
Measuring this state in the computational basis produces an ordinary
uniform random example. A quantum learner can instead apply unitary
transformations before measuring, allowing the labeled examples in the
superposition to interfere. The quantum learners considered in this paper  use copies of this state and make no
quantum membership queries.\footnote{A quantum membership oracle provides the unitary
transformation
$
    O_f|x,b\rangle
    =
    |x,b\oplus f(x)\rangle.
$
A quantum membership query can be used to generate a quantum random example under the uniform distribution,
but a quantum random example does not in general allow the
learner to implement $O_f$ on arbitrary superpositions of its choice.}\\

As usual, a quantum learning algorithm is described by a uniform sequence $\{Q_n\}_{n \geq 1}$ of quantum circuits. When we say that a quantum learner runs in time $T(n)$, this means that each circuit $Q_n$ contains at most $T(n)$ gates, and given $n$ we can print the description of $Q_n$ in deterministic time $T(n)$.

\paragraph{Quantum Notation.} We write $\Had$ for the single-qubit Hadamard transform, so that
the letter $H$ remains reserved for sign-valued hypotheses (\Cref{sec:basic_notation}). In our notation, for each $x$,
$$
    \Had|f(x)\rangle
    =
    \frac1{\sqrt2}
    \left(
        |0\rangle+(-1)^{f(x)}|1\rangle
    \right)
    =
    \frac1{\sqrt2}\left(|0\rangle+F(x)|1\rangle\right).
$$

Every quantum learning algorithm described in this paper produces a classical deterministic Boolean circuit as its output hypothesis.

\subsection{Probability}

The following inequality will be helpful during the analysis of the success probability of the quantum learner.

\paragraph{The Paley-Zygmund Inequality.} This is a standard second moment estimate that converts moment bounds into a lower bound on the probability that a 
random variable is not too small. We include here a precise formulation and its short proof for completeness.

\begin{lemma}[Paley-Zygmund Inequality]\label{lem:PZ-Inequality}
Let $X$ be a random variable satisfying
$$
0<\E[X]<\infty
\qquad\text{and}\qquad
0 < \E[X^2]<\infty.
$$
Then, for every $\lambda\in[0,1)$,
$$
\Pr\!\left[X\geq\lambda \cdot \E[X]\right]
\geq
(1-\lambda)^2 \cdot 
\frac{\E[X]^2}{\E[X^2]}.
$$
\end{lemma}

\begin{proof}
Consider the event
$$
E_\lambda
\eqdef
\{X\geq\lambda \cdot \E[X]\}.
$$
Since $X<\lambda\E[X]$ on the complementary event $E_\lambda^{\mathsf c}$,
\begin{equation}\nonumber
\E[X] \;=\;  \E\!\left[X\mathbf{1}_{E_\lambda^{\mathsf c}}\right]
+
\E\!\left[X\mathbf{1}_{E_\lambda}\right] 
 \;\leq\;
\lambda\E[X]
+
\E\!\left[X\mathbf{1}_{E_\lambda}\right].
\end{equation}
Consequently,
\begin{equation}\nonumber
(1-\lambda)\E[X]
\leq
\E\!\left[X\mathbf{1}_{E_\lambda}\right].
\end{equation}
By the Cauchy-Schwarz inequality,
$$
\E\!\left[X\mathbf{1}_{E_\lambda}\right]
\leq
\E[X^2]^{1/2}
\E\!\left[\mathbf{1}_{E_\lambda}^2\right]^{1/2}
=
\E[X^2]^{1/2}
\Pr[E_\lambda]^{1/2}.
$$
Combining these inequalities and squaring both sides gives the desired inequality
\begin{equation*}
\Pr[E_\lambda]
\geq
(1-\lambda)^2 \cdot 
\frac{\E[X]^2}{\E[X^2]}.
\qedhere
\end{equation*}
\end{proof}

\subsection{$4$-Wise Independence}\label{sec:kWiseIndependentFamilies}

We will need the following definition, which will be used later with $k = 4$.

\paragraph{$k$-Wise Independence.} Let $\mathcal X$ be a (nonempty) finite domain, let $\mathcal A$ be a finite seed set, and let
$$
\mathcal H
=
\{h_a \colon\mathcal X\to\bits\}_{a\in\mathcal A}
$$
be a family of Boolean functions indexed by elements in $\mathcal A$. For an integer $k\geq1$, the family $\mathcal H$ is called \emph{$k$-wise independent} if, for every $1\leq\ell\leq k$, every set of distinct points $x^{(1)},\ldots,x^{(\ell)}\in\mathcal X$, and every vector $v\in\bits^\ell$,
$$
\Pr_{a \sim A}\left[
\bigl(
h_{a}(x^{(1)}),\ldots,h_{a}(x^{(\ell)})
\bigr)
=v
\right]
=
2^{-\ell}.
$$
Equivalently, the evaluations at any $\ell\leq k$ distinct inputs are independent uniform bits.

For each $h_a$, define its sign representation
$
H_a(x)
=
(-1)^{h_a(x)}.
$
Since $b\mapsto(-1)^b$ is a bijection from $\bits$ to $\pmone$, the family $\mathcal H$ is $k$-wise independent precisely when
$
\bigl(
H_{\bm{a}}(x^{(1)}),\ldots,H_{\bm{a}}(x^{(\ell)})
\bigr),
$
for $\bm{a} \sim A$, is uniform on $\pmone^\ell$ for every collection of $\ell \leq k$ distinct inputs.

Similarly, we will say that a collection of random variables $\{Y_x\}_{x \in \mathcal{X}}$ is $k$-wise independent if every tuple containing $\ell \leq k$ distinct random variables is uniformly distributed.

\paragraph{$4$-Wise Independence Bounds.} We next record some consequences of $4$-wise independence that will be used later. Let $N=|\mathcal{X}|$, fix an arbitrary function
$$
F\colon \mathcal{X}\to\pmone,
$$
and define, for a uniformly random seed $a \sim A$, 
$$
Y_x
\eqdef
F(x)H_{a}(x) \in \{-1,1\},
\qquad
Z_{a}
\eqdef
\frac{1}{\sqrt N}
\sum_{x\in\mathcal{X}}Y_x.
$$
Equivalently, the random variable
$$
Z_{\bm{a}}
=
\sqrt N \cdot 
\E_{x\sim U_n}[F(x)H_{\bm{a}}(x)]
$$
is the normalized correlation between the target function $F$ and the randomly selected hypothesis $H_{\bm{a}}$.

\begin{lemma}[Moment Bounds for Correlation with $4$-Wise Independent Family]\label{lem:fourWiseMomentBounds}
If the family $\{H_a\}_{a\in\mathcal A}$ is $4$-wise independent, then for every fixed target $F \colon \mathcal{X}\to\pmone$, the random variable $Z_{\bm{a}}$ defined above satisfies
$$
\E_{a \sim A}[Z_a]=0,
\qquad
\E_{a \sim A}[Z_a^2]=1,
\qquad
\E_{a \sim A}[Z_a^4]
=
3-\frac{2}{N}
\leq3.
$$
\end{lemma}

\begin{proof} 
Multiplication by the fixed sign $F(x)$ merely exchanges the two possible values of $H_{\bm{a}}(x)$. Consequently, $
\{Y_x\}_{x\in\mathcal{X}}
$
is also $4$-wise independent.

The first moment is
$$
\E[Z_{\bm{a}}]
=
\frac{1}{\sqrt N}
\sum_x\E_{\bm{a}}[Y_x]
=
0.
$$

For the second moment,
$$
\begin{aligned}
\E_{\bm{a}}[Z_{\bm{a}}^2]
&=
\frac{1}{N}
\sum_{x_1,x_2}\E_{\bm{a}}[Y_{x_1} Y_{x_2}] \\
&=
\frac{1}{N}
\left(
\sum_x\E_{\bm{a}}[Y_x^2]
+
\sum_{x_1\neq x_2}\E_{\bm{a}}[Y_{x_1} Y_{x_2}]
\right) \\
&=
\frac{1}{N}(N+0)
=
1.
\end{aligned}
$$
Here $Y_x^2=1$, while the mixed terms vanish by pairwise independence.

For the fourth moment, 
$$
\E_{\bm{a}}[Z_{\bm{a}}^4]
=
\frac{1}{N^2}
\sum_{x_1,x_2,x_3,x_4}
\E_{\bm{a}}[
Y_{x_1}Y_{x_2}Y_{x_3}Y_{x_4}
].
$$
A term of this sum has nonzero expectation only if every input occurs an even number of times. There are $N$ tuples in which all four indices are equal. There are also\footnote{There are
$\binom N2$ choices of the unordered pair $\{x,y\}$.  Once this pair is fixed,
an ordered tuple $(x_1,x_2,x_3,x_4)$ is determined by choosing the two
positions occupied by $x$; the remaining two positions are occupied by $y$.
Thus the number of tuples is precisely
$$
    \binom N2 \binom42
    =
    6\binom N2
    =
    3N(N-1).
$$}
$3N(N-1)$
tuples containing two distinct inputs, each appearing twice. Every such tuple contributes $1$. All other tuples contribute zero by $4$-wise independence. Therefore
\begin{equation*}
\E_{\bm{a}}[Z_{\bm{a}}^4]
=
\frac{N+3N(N-1)}{N^2} 
=
3-\frac{2}{N}
\leq3.
\qedhere
\end{equation*}
\end{proof}

\paragraph{The Global Sign Flip Property.} In our learning algorithm we will require symmetry between positive and negative correlations. We say that the family $\mathcal H$ admits a \emph{global sign flip} if there is a bijection
$$
\iota \colon\mathcal A\to\mathcal A
$$
such that
$$
H_{\iota(a)}(x)
=
-H_a(x)
$$
for every seed $a$ and every input $x$. Since $\iota$ is a bijection, it preserves the uniform distribution on the seed set. Moreover,
$$
Z_{\iota(a)}
=
-Z_a,
$$
so the distribution of $Z_{\mathcal A}$ is symmetric about zero. For a $4$-wise independent family, combining this symmetry with \Cref{lem:fourWiseMomentBounds} and applying the Paley-Zygmund inequality (\Cref{lem:PZ-Inequality}) with $X = Z_{\mathcal{A}}^2$ and $\lambda = 1/4$ gives the following probability bounds:
\begin{equation}\label{eq:prob_bounds} 
\Pr_{a \sim \mathcal{A}}\left[
|Z_a|\geq\frac12
\right]
\geq
\frac{3}{16} \quad \text{and} \quad 
\Pr_{a \sim \mathcal{A}}\left[
Z_a\geq\frac12
\right]
=
\frac12
\Pr_{a \sim \mathcal{A}}\left[
|Z_a|\geq\frac12
\right]
\geq
\frac{3}{32}.
\end{equation}

\subsection{Cryptography}\label{sec:crypto}

We will need the following standard notion from cryptography. We refer to \citep{DBLP:books/cu/Goldreich2001} for more background on one-way functions.

\paragraph{Subexponentially Secure One-Way Functions.} A uniform family of functions $f_n \colon \{0,1\}^n \to \{0,1\}^{m(n)}$ computable in deterministic polynomial time is a \emph{subexponentially secure one-way function} if there is a constant $\varepsilon > 0$ such that, for every uniform randomized  algorithm $\mathcal{I}$ running in time at most $2^{n^\varepsilon}$, every positive polynomial $p$, and every sufficiently large $n$,
$$
\Pr_{x \sim \mathcal{U}_n,\;y = f_n(x),\;\mathcal{I}}[\mathcal{I}(1^n, y) \in f^{-1}_n(y)] < \frac{1}{p(n)}.
$$

\section{Quantum Learning Algorithms: Proof of \Cref{thm:quantum_learners}}

The proof follows the outline presented in \Cref{sec:techniques}. We first describe a classical learner with a weaker guarantee on its success probability. We then explain how to improve it beyond $1/2$ using a quantum algorithm. Both constructions rely on a suitable $4$-wise independent function family, which we describe next. 

\subsection{Technical Background}

\subsubsection{Quadratic Functions: A Suitable 4-Wise Independent Family}\label{sec:four-wise-family}

As in \Cref{sec:kWiseIndependentFamilies}, let $\mathcal X$ be a finite domain with $N = 2^n =  |\mathcal X|$, let $\mathcal A$ be a finite seed set, and let
$$
\mathcal H
\eqdef
\{h_a \colon\mathcal X\to\bits\}_{a\in\mathcal A}
$$
be a family of Boolean functions indexed by elements in $\mathcal A$. Recall that for each $h_a$, we define its \emph{sign representation}
$
H_a(x)
=
(-1)^{h_a(x)}.
$

\paragraph{The Quadratic Boolean Family.} We now construct an explicit family having all required properties. Let
$$
r
\eqdef 
1+n+\binom n2.
$$
A seed $a\in\bits^r$ is interpreted as
$$
a
=
\left(
a_{\emptyset},
\{a_i\}_{i=1}^n,
\{a_{ij}\}_{1\leq i<j\leq n}
\right).
$$
Associate with this seed the multilinear quadratic polynomial
$$
q_a(x)
=
a_{\emptyset}
\oplus
\bigoplus_{i=1}^n a_ix_i
\oplus
\bigoplus_{1\leq i<j\leq n}a_{ij}x_ix_j,
$$
where all arithmetic is over $\F_2$. Finally, define
$$
h_a(x)
=
q_a(x),
\qquad
H_a(x)
=
(-1)^{q_a(x)}.
$$

We will need the following result about the $4$-wise independence of quadratic Boolean functions.\footnote{For a more general version of this construction, we refer to~\citep{DBLP:conf/soda/CalderbankGLMS05}.}

\begin{lemma}[Properties of the Quadratic Boolean Family]\label{lem:prop_family}
For every $n\geq1$, the family
$$
\mathcal H_n
=
\{H_a:\bits^n\to\pmone\}_{a\in\bits^r}
$$
has the following properties.

\begin{enumerate}
\item
The family is $4$-wise independent under a uniform choice of the seed $a$.

\item
The family admits a global sign flip $\iota:\mathcal A\to\mathcal A$.

\item
There is an injective map
$$
J:\{\perp\}\cup\bits^n\to\bits^r
$$
such that $J(\perp)=0^r$ and
$$
H_a(x)
=
(-1)^{a\cdot J(x)}.
$$
Both $J$ and its inverse on its image are efficiently computable.

\item
For every seed $a$, the Boolean function $h_a$ is computable by a deterministic classical Boolean circuit of size $O(n^2)$.
\end{enumerate}
\end{lemma}

\begin{proof}
We provide a detailed proof for completeness. Define the feature map
$$
J(\perp)
=
0^r,
\qquad
J(x)
=
\left(
1,
\{x_i\}_{i=1}^n,
\{x_ix_j\}_{1\leq i<j\leq n}
\right).
$$
The ordinary binary inner product satisfies
$
a\cdot J(x)
=
q_a(x)
$
over $\F_2$, and hence
$
H_a(x)
=
(-1)^{a\cdot J(x)}.
$

To prove $4$-wise independence, fix distinct points
$$
x^{(1)},\ldots,x^{(k)}\in\bits^n,
\qquad
1\leq k\leq4.
$$
Their output bits are obtained from the seed through the linear map
$$
\begin{aligned}
L:\F_2^r&\longrightarrow\F_2^k, \\
a&\longmapsto
\left(
a\cdot J(x^{(1)}),\ldots,
a\cdot J(x^{(k)})
\right).
\end{aligned}
$$
We argue next that the vectors
$$
J(x^{(1)}),\ldots,J(x^{(k)})
$$
are linearly independent over $\F_2$. Suppose, toward a contradiction, that these feature vectors are linearly dependent. Since the coefficients in a binary linear dependence belong to $\bits$, there is a nonempty set
$$
T\subseteq\{x^{(1)},\ldots,x^{(k)}\}
$$
such that
$$
\sum_{x\in T}J(x)=0.
$$
The constant coordinate of every $J(x)$ is $1$, so the constant coordinate of this sum is $|T|$ modulo $2$. Thus $|T|$ must be even. Since $1\leq|T|\leq4$, it remains to consider the cases $|T|=2$ and $|T|=4$.

If $T=\{x,y\}$, the linear coordinates of $J(x)+J(y)=0$ imply
$
x+y=0,
$
and hence $x=y$. This contradicts the assumption that the selected points are distinct.

Suppose now that
$$
T
=
\{x^{(1)},x^{(2)},x^{(3)},x^{(4)}\}.
$$
The linear coordinates of the assumed relation give
$$
x^{(1)}+x^{(2)}+x^{(3)}+x^{(4)}
=
0.
$$
Set
$$
u=x^{(1)},
\qquad
v=x^{(2)}+x^{(1)},
\qquad
w=x^{(3)}+x^{(1)}.
$$
Then the four points are
$$
u,
\qquad
u+v,
\qquad
u+w,
\qquad
u+v+w.
$$
Because the four points are distinct, $v$ and $w$ are nonzero and distinct. Two nonzero distinct vectors over $\F_2$ are linearly independent. The matrix whose rows are $v$ and $w$ therefore has rank $2$, so it has a nonzero $2$ by $2$ minor. Consequently, there are coordinates $i<j$ such that
$$
v_iw_j+v_jw_i
=
1.
$$
For these coordinates, the sum of the quadratic feature $x_ix_j$ over the four points is
$$
\begin{aligned}
&u_iu_j
+(u_i+v_i)(u_j+v_j)
+(u_i+w_i)(u_j+w_j) \\
&\qquad
+(u_i+v_i+w_i)(u_j+v_j+w_j) \\
&=
v_iw_j+v_jw_i \\
&=
1.
\end{aligned}
$$
This contradicts the assumption that every coordinate of
$
\sum_{x\in T}J(x)
$
is zero. Hence every collection of at most four distinct feature vectors is linearly independent.

From linear independence, it follows that $L$ has rank $k$. The rank nullity theorem gives
$\dim(\ker L)=r-k$, and a standard linear algebra argument shows that every element of $\F_2^k$ has exactly $2^{r-k}$ preimages under $L$. A uniform seed $a\in\F_2^r$ consequently produces a uniform evaluation vector
$$
\bigl(
h_a(x^{(1)}),\ldots,h_a(x^{(k)})
\bigr)
\in\F_2^k.
$$
This proves $4$-wise independence.

For the global sign flip, define $\iota$ by replacing the constant coefficient $a_{\emptyset}$ with $a_{\emptyset}\oplus1$ and leaving every other coefficient unchanged. Then
$
q_{\iota(a)}(x)
=
q_a(x)\oplus1
$
for every $x$, and therefore
$
H_{\iota(a)}(x)
=
-H_a(x).
$

Note that $J$ is injective on $\{\perp\}\cup\bits^n$. Its constant coordinate separates $J(\perp)=0^r$ from every $J(x)$, while its linear coordinates recover $x$ and therefore distinguish any two Boolean inputs. Its inverse on its image returns $\perp$ on $0^r$ and otherwise reads $x$ from the linear coordinates. 

Finally, computing $J(x)$, evaluating $q_a(x)$, and implementing the corresponding hypothesis circuit require $O(n^2)$ elementary Boolean operations.
\end{proof}

\subsubsection{Derandomized Classical Learner with $p^\star \leq 1/2$}

In this section, we describe a zero-query polynomial-time classical learner in the large error regime that succeeds with  probability $\Omega(1)$. The quantum learner presented in \Cref{sec:quantum_learner} builds on the construction described here.

\begin{proposition}[Classical Zero-Query Learner]\label{prop:zero-query-construction}
There is a randomized zero-query algorithm $\mathcal{R}$ that runs in $\poly(n)$ time and outputs a deterministic classical Boolean circuit $h$ such that, for every target $f \colon\bits^n\to\bits$,

\[
    \Pr_{h \leftarrow \mathcal{R}}\left[
        \Pr_x[h(x)=f(x)]\ge \frac12+\frac14\,2^{-n/2}
    \right]
    \ge \frac{3}{32}.
\]
\end{proposition}

\begin{proof}
We use the efficiently computable $4$-wise independent family $\mathcal{H}$ from \Cref{sec:four-wise-family}. The learner $\mathcal{R}$ samples a uniform seed $a$ and outputs the corresponding deterministic circuit $h_a$, whose sign representation is $H_a$. By \Cref{lem:prop_family}, $\mathcal{R}$ runs in polynomial time and its output hypothesis is a deterministic Boolean circuit.

For a fixed target $f \colon \{0,1\}^n \to \{0,1\}$ with sign representation $F(x) = (-1)^{f(x)}$, set
\[
    Z_a\eqdef\frac1{\sqrt N}\sum_xF(x)H_a(x).
\]
All probabilities and expectations in the remainder of this proof are over
the uniformly sampled seed $a$. The target $F$ remains fixed. By \Cref{lem:fourWiseMomentBounds},
\[
    \E [Z_a]=0,
    \qquad
    \E [Z_a^2]=1,
    \qquad
    \E [Z_a^4]\le 3.
\]
Recall that the family $\mathcal{H}$ admits a global sign flip (\Cref{lem:prop_family}). Consequently, applying the Paley-Zygmund Inequality (\Cref{lem:PZ-Inequality}) as in the derivation of \Cref{eq:prob_bounds} from \Cref{sec:kWiseIndependentFamilies}, we get that
\begin{equation}\nonumber
   \Pr\! \left[
Z_a\geq\frac12
\right]
\geq
\frac{3}{32}.
\end{equation}
On this event, $\corr(H_a,F)=Z_a/\sqrt N\ge(1/2)\cdot (1/\sqrt N)$, which provides the guarantee in the statement of \Cref{prop:zero-query-construction} via \Cref{eq:agreement_vs_correlation}. 
\end{proof}

\subsection{The Correlation-Biased Quantum Sampler: Proof of Item 1}\label{sec:quantum_learner}

In this section, we describe a quantum learning algorithm that uses a single quantum random example and achieves success probability strictly larger than $1/2$. Note that the
learning procedure is quantum, but the hypothesis it returns is classical.

\subsubsection{Hypothesis Family $\mathcal{H}$ and Isometry $W$}
\label{sec:LearningTemplate}

We will need the following notions to describe the quantum algorithm $\mathcal Q$ from \Cref{thm:quantum_learners}.

\paragraph{The Hypothesis Family $\mathcal{H}$.}
Let $\mathcal H=\{H_a \colon\bits^n\to\pmone\}_{a\in\mathcal A}$ be an efficiently computable $4$-wise independent family of sign-valued functions. As in \Cref{lem:prop_family}, we assume that the seed set $\mathcal{A}$ is identified  with $\bits^r$. In particular, $$M \eqdef|\mathcal A|=2^r,$$ and expressions such as $a\cdot J(x)$ below are ordinary binary inner products in this chosen encoding.  We also assume a \emph{global sign flip}: there is a bijection $\iota \colon \mathcal A\to\mathcal A$ preserving the uniform seed distribution such that
\[
    H_{\iota(a)}(x)=-H_a(x) \quad\text{for every }a\in\mathcal A\text{ and every }x.
\]
Finally, assume that there is an efficiently computable injective feature map
\[
    J \colon\{\perp\}\cup\bits^n\to\bits^r,
\]
whose inverse on its image is also efficiently computable.  We require
\[
   J(\perp)=0^r \quad \text{and} \quad H_a(x)=(-1)^{a\cdot J(x)}.
\]

\paragraph{The Isometry $W$.} We now introduce a map $W$ needed by the quantum algorithm. Let $\perp\,\notin\bits^n$ be an auxiliary reference index. We use the
following $N+1$ basis states of an $n+1$ qubit index register:
$$
    |\mathord{\perp}\rangle \eqdef|0\rangle|0^n\rangle,
    \qquad
    |x\rangle_{\mathrm{idx}}\eqdef|1\rangle|x\rangle
    \quad (x\in\bits^n).
$$
Writing $|\ell\rangle$ for the encoded state of an index
$\ell\in\{\perp\}\cup\bits^n$, we define
$$W|\ell\rangle\eqdef\Had^{\otimes r}|J(\ell)\rangle.$$ The standard Hadamard identity
$$
    \Had^{\otimes r}|y\rangle
    =
    \frac{1}{\sqrt M}
    \sum_{a\in\bits^r}(-1)^{a\cdot y}|a\rangle
$$
and the defining properties of $J$ give
\begin{equation}\label{eq:mapW}
     W|\mathord{\perp}\rangle=\frac1{\sqrt M}\sum_{a\in\mathcal A}|a\rangle,
    \qquad
    W|x\rangle_{\mathrm{idx}}=\frac1{\sqrt M}\sum_{a\in\mathcal A}H_a(x)|a\rangle.
\end{equation}
Since $J$ is injective and the Hadamard transform is unitary, it is not hard to check that $W$ is an
isometry.\footnote{Recall that an isometry is a linear map $W$ that preserves inner products:
$\langle W\psi \mid W\phi\rangle=\langle\psi\mid\phi\rangle$
for all vectors in its domain. In particular, it  maps an orthonormal basis to an orthonormal set in the output space,
which may have larger dimension.}

 To use $W$ in our polynomial-time quantum learner, we must also
show that it can be implemented using polynomially many quantum gates.
For completeness, we describe this implementation next. Using a fresh $r$ qubit register, reversibly compute
$$
    |\ell\rangle|0^r\rangle
    \longmapsto
    |\ell\rangle|J(\ell)\rangle.
$$
Next, evaluate $J^{-1}$ on the second register and XOR the recovered
binary encoding of $\ell$ into the first register, obtaining
$$
    |\ell\rangle|J(\ell)\rangle
    \longmapsto
    |0^{n+1}\rangle|J(\ell)\rangle.
$$
This resets the original index register while retaining its information
in the second register. Temporary work qubits are returned to zero after each
evaluation. Finally, apply $\Had^{\otimes r}$ to the second register:
$$
    |0^{n+1}\rangle|J(\ell)\rangle
    \longmapsto
    |0^{n+1}\rangle W|\ell\rangle.
$$
The first register is now in a fixed zero state and is unentangled with
the second, so we can ignore it. By linearity, the second register
contains $W|\psi\rangle$ for every input state $|\psi\rangle$ in the
domain of $W$. (This is the seed register measured by the learner.) The assumed efficient evaluation and inversion of $J$
admit reversible implementations with polynomial overhead. Together
with $r=\poly(n)$ for the concrete family, this gives an implementation
using $\poly(n)$ gates.

\subsubsection{Quantum Algorithm $\mathcal{Q}$}
\label{sec:tiltedSamplerAlgorithm}

Using the concepts and notation introduced above, the quantum learner can be described in  four steps. As usual, let $N = 2^n$, and let $f \colon \{0,1\}^n \to \{0,1\}$ be the target function. Moreover, let $\mathcal{H}$, $W$, $\mathcal{A}$, and $M = |\mathcal{A}| = 2^r$ be as in \Cref{sec:LearningTemplate}.\\

\noindent \textbf{Quantum Procedure $\mathcal Q$:}

\begin{enumerate}[label=\arabic*.]
\item
Receive one uniform quantum random example
$$
    |\mathrm{QEX}_f\rangle
    =
    \frac1{\sqrt N}\sum_x|x\rangle|f(x)\rangle.
$$
Apply a Hadamard gate to the label qubit.  Using
$F(x)=(-1)^{f(x)}$ and linearity, 
$$
\begin{aligned}
    |\mathrm{QEX}_f\rangle
    &\longmapsto
    \frac1{\sqrt{2N}}\sum_x
        |x\rangle\bigl(|0\rangle+F(x)|1\rangle\bigr) \\
    &=
    \frac1{\sqrt{2N}}\sum_x|x\rangle|0\rangle
    +
    \frac1{\sqrt{2N}}\sum_xF(x)|x\rangle|1\rangle.
\end{aligned}
$$
Regard the rightmost qubit as a flag and place it first in the notation.  The same
state can be rewritten as
$$
    \frac1{\sqrt2}|0\rangle
        \left(\frac1{\sqrt N}\sum_x|x\rangle\right)
    +
    \frac1{\sqrt2}|1\rangle
        \left(\frac1{\sqrt N}\sum_xF(x)|x\rangle\right).
$$
(The flag $0$ branch is a uniform state, while the flag $1$ branch
contains the target values as phases. The flag is left unmeasured, allowing the ``reference'' branch and the ``target'' branch to interfere in Step 3.)

\item
Apply $\Had^{\otimes n}$ to the input register only on the flag $0$ branch.
Because
$$
    \Had^{\otimes n}
    \left(\frac1{\sqrt N}\sum_x|x\rangle\right)
    =
    |0^n\rangle,
$$
the resulting state can be written as
$$
    \frac1{\sqrt2}|0\rangle|0^n\rangle
    +
    \frac1{\sqrt{2N}}\sum_xF(x)|1\rangle|x\rangle.
$$
As in \Cref{sec:LearningTemplate}, encode the reference label by
$|\mathord{\perp}\rangle=|0\rangle|0^n\rangle$ and encode an input label $x$
by $|x\rangle_{\rm lab}=|1\rangle|x\rangle$.  In this label space the state is
$$
    |\psi_F\rangle
    \eqdef
    \frac1{\sqrt2}|\mathord{\perp}\rangle
    +
    \frac1{\sqrt{2N}}\sum_xF(x)|x\rangle_{\rm lab}.
$$
(Note that this preparation is exact.  It uses the given quantum example, one Hadamard
gate, and $n$ Hadamard gates controlled on the flag being $0$.  It makes no membership query and
uses no measurement or postselection.)

\item
Apply the isometry $W$.  By \Cref{eq:mapW} and  linearity,
$$
\begin{aligned}
    W|\psi_F\rangle
    &=
    \frac1{\sqrt2}W|\mathord{\perp}\rangle
    +
    \frac1{\sqrt{2N}}\sum_xF(x)W|x\rangle_{\rm lab} \\
    &=
    \frac1{\sqrt{2M}}\sum_a|a\rangle
    +
    \frac1{\sqrt{2NM}}\sum_x\sum_aF(x)H_a(x)|a\rangle \\
    &=
    \frac1{\sqrt{2M}}\sum_{a\in\mathcal A}(1+Z_a)|a\rangle.
\end{aligned}
$$
where
$$
    Z_a=\frac1{\sqrt N}\sum_x F(x)H_a(x)
        =\sqrt N\,\corr(H_a,F).
$$
(The amplitude of the basis state $|a\rangle$ is therefore proportional to
$1+Z_a$.  In this sense, the normalized correlation between $H_a$ and $F$ has
been encoded into a quantum amplitude.)

\item
Measure the seed register.  The probability of observing a basis state is the
squared absolute value of its amplitude, so the resulting seed is distributed
according to
\begin{equation}\label{eq:prob_dist}
    \mathsf P_F(a) \eqdef \frac{(1+Z_a)^2}{2M}.      
\end{equation}
Once $a$ has been measured, output the corresponding deterministic classical
 circuit $h_a$. 
\end{enumerate}

\paragraph{Intuition.} It is helpful to isolate the role of the reference branch.  Without
$|\mathord{\perp}\rangle$, the amplitude of $|a\rangle$ would be proportional to $Z_a$,
so measurement would sample $a$ with probability proportional to $Z_a^2$.
Squaring would erase the sign of the correlation, just as in ordinary Fourier
sampling (\Cref{sec:fourier_sampling}). In contrast, the reference branch adds the constant amplitude $1$, producing
$$
    (1+Z_a)^2=1+2Z_a+Z_a^2.
$$
The linear term $2Z_a$ favors positive correlations over negative ones.  More
concretely, the global sign flip pairs a seed with normalized correlation $z$
with a seed having normalized correlation $-z$.  For $z>0$, their measurement
weights differ by
$$
    (1+z)^2-(1-z)^2=4z>0.
$$
Interference changes the sampling distribution before measurement so that the
positively correlated orientation is intrinsically more likely.

\subsubsection{Correctness}
\label{sec:oneExampleGuarantee}

We now prove that the quantum procedure $\mathcal Q$ from
\Cref{sec:tiltedSamplerAlgorithm} outputs a positively correlated
hypothesis with probability strictly greater than $1/2$. This completes the proof of \Cref{thm:quantum_learners} Item (i).

\begin{theorem}[Correctness of the Quantum Learner $\mathcal{Q}$] \label{thm:quantum_correctness}
For every $n\geq1$ and every target function $f\colon\bits^n\to\bits$, $\mathcal Q$ uses one uniform quantum random example and runs in time $\poly(n)$. With probability
$p^\star \geq 0.53$, it outputs a deterministic classical Boolean
circuit $h$ such that
$$
    \Pr_{x\sim\mathcal U_n}[h(x)=f(x)]
    \geq \frac12+ a_0 \cdot 2^{-n/2},
$$
where $a_0 = 0.001$.
\end{theorem}

\begin{proof} 
Fix the target function $F(x)=(-1)^{f(x)}$. Let $\mathsf U$ be the uniform
distribution on the seed set $\mathcal A$, and let $\mathsf P_F$
be the measurement distribution of $\mathcal Q$ (\Cref{sec:tiltedSamplerAlgorithm}). Recall that 
$Z_a \eqdef \sqrt N \cdot \corr(H_a,F)$.
We write $Z$ for this
function of the seed and indicate the seed distribution ($\mathsf U$ or $\mathsf P_F$) by the
subscript on each probability or expectation.

Set $c \eqdef 2a_0=0.002$. By
\Cref{eq:agreement_vs_correlation} and $N=2^n$, a measured seed
$a$ yields the desired agreement precisely when $Z_a\geq c$.
Our goal is therefore to show that
$$
    \Pr_{\mathsf P_F}[Z\geq c]\geq0.53.
$$

We first recall the properties of $Z$ under the uniform seed
distribution. By \Cref{lem:prop_family} and \Cref{lem:fourWiseMomentBounds},
$$
    \E_{\mathsf U}[Z^2]=1,
    \qquad
    \E_{\mathsf U}[Z^4]\leq3.
$$
The global sign flip also makes $Z$ symmetric about zero under
$\mathsf U$, as explained in \Cref{sec:kWiseIndependentFamilies}.
In particular, its odd moments vanish.

To use these facts under the quantum measurement distribution,
recall from Step 4 of \Cref{sec:tiltedSamplerAlgorithm} that
$\mathsf P_F(a)=(1+Z_a)^2/(2M)$. Since $\mathsf U(a)=1/M$,
for every function $g\colon\mathbb R\to\mathbb R$ we have
$$
    \E_{\mathsf P_F}[g(Z)]
    =
    \frac12\E_{\mathsf U}[(1+Z)^2g(Z)].
$$
This identity lets us express moments under $\mathsf P_F$ in
terms of moments under $\mathsf U$.

The classical learner in \Cref{prop:zero-query-construction}
applies Paley-Zygmund (\Cref{lem:PZ-Inequality}) under the uniform seed distribution.
Here we apply the same inequality under the quantum measurement
distribution, to the nonnegative random variable
$$
    X \eqdef \max\{Z,0\}.
$$
Since $c>0$, the events $X\geq c$ and $Z\geq c$ coincide.
We will therefore bound the first two moments of $X$ under
$\mathsf P_F$.

For the first moment, the conversion identity and the discussion above on the (uniform) moments of $Z$ give
$$
\begin{aligned}
    \E_{\mathsf P_F}[Z]
    &=\frac12\E_{\mathsf U}[Z(1+Z)^2] \\
    &=\frac12\left(
        \E_{\mathsf U}[Z]
        +2\E_{\mathsf U}[Z^2]
        +\E_{\mathsf U}[Z^3]
      \right)
    =1,
\end{aligned}
$$
where the last equality uses symmetry and
$\E_{\mathsf U}[Z^2]=1$. Since $X\geq Z$, it follows that
$$
    \mu\eqdef\E_{\mathsf P_F}[X]\geq1.
$$

For the second moment, observe that
$X^2=Z^2\1_{\{Z>0\}}$. The function
$z^2(1+|z|)^2$ is even and vanishes at zero, so symmetry under
$\mathsf U$ assigns half of its expectation to positive values
of $Z$. Applying the conversion identity and using that $Z$ is symmetric about zero under $\mathsf U$,
we obtain
$$
\begin{aligned}
    \nu\eqdef\E_{\mathsf P_F}[X^2]
    &=\frac12\E_{\mathsf U}
        [Z^2(1+Z)^2\1_{\{Z>0\}}] \\
    &=\frac14\E_{\mathsf U}[Z^2(1+|Z|)^2] \\
    &=\frac14\left(
        1+2\E_{\mathsf U}[|Z|^3]
        +\E_{\mathsf U}[Z^4]
      \right).
\end{aligned}
$$
By the Cauchy-Schwarz inequality and the uniform moment bounds,
$$
    \E_{\mathsf U}[|Z|^3]
    =\E_{\mathsf U}[|Z|\,Z^2]
    \leq
    \sqrt{\E_{\mathsf U}[Z^2]\E_{\mathsf U}[Z^4]}
    \leq\sqrt3.
$$
Consequently,
$$
    \nu\leq\frac{1+2\sqrt3+3}{4}
    =1+\frac{\sqrt3}{2}.
$$

We can now apply Paley-Zygmund. Since $\mu\geq1$ and $c=0.002$,
the parameter $\lambda \eqdef c/\mu$ belongs to $(0,1)$.
Applying \Cref{lem:PZ-Inequality} to $X$ under $\mathsf P_F$
with this parameter gives
$$
\begin{aligned}
    \Pr_{\mathsf P_F}[Z\geq c]
    &=\Pr_{\mathsf P_F}[X\geq c] = \Pr_{\mathsf P_F}[X \geq \lambda \cdot \mu] \\
    &\geq
      \left(1-\frac{c}{\mu}\right)^2
      \frac{\mu^2}{\nu} 
    =\frac{(\mu-c)^2}{\nu} \\
    &\geq\frac{(1-c)^2}{1+\sqrt3/2} 
    =\frac{(0.998)^2}{1+\sqrt3/2} \\
    & >0.53.
\end{aligned}
$$
As noted at the beginning of the proof, this establishes the
required agreement guarantee.

Finally, $\mathcal{Q}$ uses one uniform
quantum random example. The state preparation in Steps 1 and 2
and the implementation of $W$ in \Cref{sec:LearningTemplate}
give the polynomial running time. By \Cref{lem:prop_family},
the measured seed specifies a deterministic classical Boolean
circuit of size $O(n^2)$. This proves the theorem.
\end{proof}

\subsection{Boosting Correlation and Success Probability: Proof of Item 2}
\label{sec:majorityAmplification}

We now prove Item (ii) of \Cref{thm:quantum_learners}. The quantum algorithm $\mathcal Q^+$ runs
$\mathcal Q$ independently on several quantum random examples and takes
the majority of the resulting hypotheses. The key property is that, at
each input, a sampled hypothesis has a small positive bias toward the
target. Moreover, these agreement indicators are independent
across distinct input pairs. Majority increases the pointwise bias, and 
independence among pairs then gives concentration of the overall correlation. We provide the details next.

Fix a target $f\colon\bits^n\to\bits$, write $F=(-1)^f$, and let
$N=2^n$. We use the quadratic Boolean family from
\Cref{lem:prop_family}. Let $\mathsf U$ be the uniform distribution on
its seed set $\mathcal A$, let $M=|\mathcal A|$, and recall that the
measurement distribution of $\mathcal Q$ is
$$
    \mathsf P_F(a)=\frac{(1+Z_a)^2}{2M},
    ~~ \text{where}~~
    Z_a=\frac1{\sqrt N}\sum_x F(x)H_a(x).
$$

We will need the following lemmas to prove the correctness of $\mathcal{Q}^+$.

\begin{lemma}[Pointwise Bias and  Independence]
\label{lem:pointwise-bias}
For $a\sim\mathsf P_F$, define $Y_x\eqdef F(x)H_a(x)$. The following statements hold:
\begin{itemize}
    \item $\E_{\mathsf P_F}[Y_x]=\frac1{\sqrt N}$ for every $x$.
    \item    
    $\E_{\mathsf P_F}[Y_xY_y]=\frac1N
    $ for every $x\ne y$.
    \item For $x \neq y$, the random variables $Y_x$ and $Y_y$ are independent under $\mathsf P_F$.
\end{itemize}
\end{lemma}

\begin{proof}
Regard $Y_x$ and $Z=Z_a$ as functions of the seed. As above, under $\mathsf U$,
the signs $Y_x$ are unbiased and $4$-wise independent. Similarly to the proof of \Cref{thm:quantum_correctness}, for any function
$G$ of the seed, the change of distribution gives
\begin{equation}\label{eq:transfer_dist}
        \E_{\mathsf P_F}[G]
    =\frac12\E_{\mathsf U}[(1+2Z+Z^2)G].
\end{equation}

For a fixed input $x$, the uniform moments needed here are
$$
    \E_{\mathsf U}[Y_x]=0,
    \qquad
    \E_{\mathsf U}[ZY_x]=\frac1{\sqrt N},
    \qquad
    \E_{\mathsf U}[Z^2Y_x]=0.
$$
Indeed, in
$\E_{\mathsf U}[ZY_x]=N^{-1/2}\sum_u\E_{\mathsf U}[Y_uY_x]$,
only $u=x$ contributes. In the expansion of $Z^2Y_x$, cancellation
of repeated factors leaves either one unbiased sign or three distinct
unbiased signs, so every term has expectation zero due to the $4$-wise independence of $\mathcal{H}$. Consequently, \Cref{eq:transfer_dist} with $G = Y_x$ yields
$$
    \E_{\mathsf P_F}[Y_x]
    =\frac12\left(0+\frac2{\sqrt N}+0\right)
    =\frac1{\sqrt N}.
$$

Now fix distinct inputs $x,y$. We will rely on  \Cref{eq:transfer_dist} again. The same cancellation argument gives
$$
    \E_{\mathsf U}[Y_xY_y]=0,
    \qquad
    \E_{\mathsf U}[ZY_xY_y]=0.
$$
For the quadratic term,
$$
    \E_{\mathsf U}[Z^2Y_xY_y]
    =\frac1N\sum_{u,v}\E_{\mathsf U}[Y_uY_vY_xY_y]
    =\frac2N.
$$
Only $(u,v)=(x,y)$ and $(u,v)=(y,x)$ contribute, each with value
$1$. Every other term reduces to a product of two or four distinct
unbiased signs and vanishes by $4$-wise independence. Thus
$\E_{\mathsf P_F}[Y_xY_y]=1/N$.

For the last item, let $x \neq y$. For any sign $Y\in\{-1,1\}$ and any $s\in\{-1,1\}$, we have the pointwise identity
$$
\mathbf 1_{\{Y=s\}}=\frac{1+sY}{2}.
$$
Indeed, the right side equals $1$ when $Y=s$ and $0$ otherwise.
Write $\mu=1/\sqrt N$, and take all probabilities and expectations below
under $\mathsf P_F$. By the first two items,
$$
\mathbb E[Y_x]=\mathbb E[Y_y]=\mu,
\qquad
\mathbb E[Y_xY_y]=\mu^2.
$$
Taking expectations in the indicator identity gives
$$
\Pr[Y_x=s]=\frac{1+s\mu}{2},
\qquad
\Pr[Y_y=t]=\frac{1+t\mu}{2}.
$$
Now let $s,t\in\{-1,1\}$ be arbitrary. Applying the same identity to both signs,
$$
\begin{aligned}
\Pr[Y_x=s,Y_y=t]
&=\mathbb E\!\left[
    \mathbf 1_{\{Y_x=s\}}\mathbf 1_{\{Y_y=t\}}
  \right]\\
&=\frac14\mathbb E[(1+sY_x)(1+tY_y)]\\
&=\frac14\left(
    1+s\mathbb E[Y_x]+t\mathbb E[Y_y]
      +st\mathbb E[Y_xY_y]
  \right)\\
&=\frac14(1+s\mu+t\mu+st\mu^2)\\
&=\frac{1+s\mu}{2}\frac{1+t\mu}{2}\\
&=\Pr[Y_x=s]\Pr[Y_y=t].
\end{aligned}
$$
This establishes the required factorization for every possible pair of values,
so $Y_x$ and $Y_y$ are independent.
\end{proof}

\begin{lemma}[Majority Bias; see  {\citep[Appendix A]{DBLP:journals/cc/Watson20}}]\label{lem:majority-bias}
There is a universal constant $0 < b \leq 1$ such that, for every odd
integer $m$ with $1\le m\le N$, independent signs $B_1,\ldots,B_m \in \{-1,+1\}$
of mean $1/\sqrt N$ satisfy
$$
    \E[\,\mathsf{Sign}(B_1,\ldots,B_m)\,]\ge b\sqrt{\frac mN},
$$
where $\mathsf{Sign}(B_1, \ldots, B_m) \in \{-1,+1\}$ denotes the sign of their sum.
\end{lemma}

\medskip

Next, we provide a precise description of $\mathcal{Q}^+$.

\paragraph{The Amplified Learner $\mathcal{Q}^+(n,m)$.}
For an odd integer $m\le N$, run $\mathcal Q$ independently $m$ times,
using one fresh quantum random example each time. Let
$a_1,\ldots,a_m$ be the measured seeds. Output the Boolean circuit
$$
    h(x)=\operatorname{Maj}
             (h_{a_1}(x),\ldots,h_{a_m}(x)),
$$
where $\operatorname{Maj}(y_1, \ldots, y_m)$ is the ordinary bit majority, and all seeds are hardwired into $h$. Its sign
representation is
$$
    H(x)=\mathsf{Sign}(H_{a_1}(x),\ldots,H_{a_m}(x)).
$$

\medskip

\begin{theorem}[Guarantee for the Amplified Learner]
\label{thm:amplified-learner}
Let $b$ be the constant in \Cref{lem:majority-bias}, and set
$C_0=4/b^2$. For every target $f\in\ALL_n$ and every odd integer
$1\le m\le N$, $\mathcal{Q}^+(n,m)$ uses exactly $m$ independent
quantum random examples, runs in time $m\cdot\poly(n)$, and outputs
a deterministic classical Boolean circuit of size $m\cdot\poly(n)$
such that
$$
    \Pr_{\mathcal{Q}^+}\left[
        \Pr_{x\sim\Un}[h(x)=f(x)]
        \ge\frac12+\frac b4\sqrt{\frac mN}
    \right]
    \ge1-\frac{C_0}{m}.
$$
\end{theorem}

\begin{proof}
For each input $x$, set $X_x=F(x)H(x)$. Multiplication of all
arguments of an odd majority by the same sign multiplies its output
by that sign, so
$$
    X_x=\mathsf{Sign}(F(x)H_{a_1}(x),\ldots,F(x)H_{a_m}(x)).
$$
The arguments are independent signs of mean $1/\sqrt N$ by
\Cref{lem:pointwise-bias}. Hence by \Cref{lem:majority-bias}  their signed majority has a common mean
$$
    \beta_m\eqdef\E[X_x]\ge b\sqrt{\frac mN}.
$$
For distinct inputs $x,y$, \Cref{lem:pointwise-bias} gives independence
of $F(x)H_{a_i}(x)$ and $F(y)H_{a_i}(y)$ for each $i$. Independence
of the runs then implies independence of the two full sequences,
and therefore of $X_x$ and $X_y$.

It follows by a simple calculation that the output correlation satisfies
$$
    \E[\corr(H,F)]=\beta_m,
    \qquad
    \operatorname{Var}(\corr(H,F))
    =\frac{1-\beta_m^2}{N}\le\frac1N.
$$
Chebyshev's inequality gives
$$
    \Pr\left[\corr(H,F)<\frac{\beta_m}{2}\right]
    \le\frac4{N\beta_m^2}\le\frac{C_0}{m}.
$$
On the complementary event, \Cref{eq:agreement_vs_correlation} yields
$$
    \Pr_{x\sim\Un}[h(x)=f(x)]
    \ge\frac12+\frac{\beta_m}{4}
    \ge\frac12+\frac b4\sqrt{\frac mN}.
$$

Each run of $\mathcal Q$ takes $\poly(n)$ time and produces a
circuit of size $O(n^2)$. A majority circuit can sum the $m$ output
bits and compare the sum with $(m+1)/2$, using
$O(m\log(m+1))$ gates. Since $m\le2^n$, this is $O(mn)$.
Thus both the complete computation and the construction of its output
circuit take $m\cdot\poly(n)$ time, and the circuit has the stated
size. In particular, it is a deterministic classical circuit built
from the quadratic Boolean hypotheses in \Cref{lem:prop_family}.
\end{proof}

\paragraph{Choosing the Failure Probability.}
Set $C=C_0+1$, and suppose $\eta\in(0,1/2)$ satisfies
$\eta\ge C/N$. Choose $m$ to be the smallest odd integer at least
$C_0/\eta$. Then
$$
    \frac{C_0}{\eta}\le m<\frac{C_0}{\eta}+2
    \le\frac{C_0+1}{\eta}\le N.
$$
Thus $m=\Theta(1/\eta)$ and $C_0/m\le\eta$. Moreover,
$$
    \frac b4\sqrt{\frac mN}
    \ge\frac{b\sqrt{C_0}}{4\sqrt\eta}\,2^{-n/2}.
$$
Taking $a_1=b\sqrt{C_0}/4$ in
\Cref{thm:amplified-learner} proves Item (ii) of
\Cref{thm:quantum_learners}, including the claimed time bound
$\poly(n)/\eta$, circuit size, and number of quantum random examples.

\section{Limitations of Classical Learners: Proof of \Cref{thm:classical_limitations}}\label{sec:limits_classical}

As alluded to in \Cref{sec:techniques}, different techniques are needed to establish each item of \Cref{thm:classical_limitations}. We first discuss the unconditional query complexity lower bound for classical learners for the concept class $\mathcal{C} = \ALL$.

\subsection{Unconditional Information Theoretic Lower Bound for $\mathcal{C} = \ALL$: Proof of Item 1}

The polynomial-time classical learner in \Cref{prop:zero-query-construction} achieves
correlation $\Omega(2^{-n/2})$ with constant success probability
(without making any queries). We now show that guaranteeing success
probability strictly greater than $1/2$ at this correlation requires
$\Omega(2^{n/2})$ classical membership queries. We note that the argument applies to
adaptive queries and places no restriction on the learner's running
time or the size of its output circuit.

\begin{theorem}[Classical Query Lower Bound]
\label{thm:classical-lower}
Let $\mathcal A$ be a randomized classical learner that makes at most
$q$ membership queries to a target $f\in\ALL_n$ and outputs a
deterministic Boolean circuit $h$. For every $\gamma>q/(2N)$,
$$
    \E_{f\sim\ALL_n} \left [ \Pr_{\mathcal A}\left[
        \Pr_{x\sim\Un}[h(x)=f(x)]\ge\frac12+\gamma
    \right] \right ]\le\frac12.
$$
Consequently, $\mathcal A$ has success probability at most $1/2$ at this correlation regime for some target $f\in\ALL_n$.
\end{theorem}

\begin{proof}
Choose $f$ uniformly from $\ALL_n$, so the signs
$F(x)=(-1)^{f(x)}$ are independent and uniform.
Fix all random coins of $\mathcal A$ and condition on a feasible
complete query transcript. Let $Q$ be the set of distinct queried
points, let $k=|Q|\le q$, and let $H=(-1)^h$ be the resulting
fixed hypothesis.

Every target agreeing with the recorded answers on $Q$ produces
the same transcript. Thus, even for adaptive queries, the
unqueried signs remain independent and uniform under this
conditioning. Consequently,
$$
    S \eqdef \sum_{x\notin Q} H(x)F(x)
$$
is a sum of $N-k$ independent uniform signs, and
$$
    N \cdot \corr(H,F)
    \;=\; \sum_{x\in Q} H(x)F(x)+S
    \;\le\; k+S.
$$
By \Cref{eq:agreement_vs_correlation}, the success event
$$
    \mathcal E \eqdef
    \left\{
        \Pr_{x\sim\Un}[h(x)=f(x)]
        \ge \frac12+\gamma
    \right\}
$$
therefore implies
$$
    S \;\ge\; 2\gamma N-k \;\ge\; 2\gamma N-q \;>\; 0,
$$
where the last inequality relied on the assumption that $\gamma>q/(2N)$. Since the conditional distribution of $S$ is symmetric about zero,
its probability of being strictly positive is at most $1/2$.
Hence the conditional probability of $\mathcal E$ is at most $1/2$.

Averaging over transcripts and random coins gives
$$
    \E_{f\sim\ALL_n}
    \left[\Pr_{\mathcal A}[\,\mathcal E\,]\right]
    \;\le\; \frac12.
$$
In particular, at least one target function has success probability no greater than
the expected value.
\end{proof}

This proves Item (i) of \Cref{thm:classical_limitations}. In particular,
for $\gamma=c \cdot 2^{-n/2}$ with fixed $c>0$, the condition is
$q<2c 2^{n/2}$, which rules out classical learners making $q=o(2^{n/2})$ queries.

\subsection{Conditional Computational Lower Bound for $\mathcal{C} = \SIZE[\poly]$: Proof of Item 2}
\label{sec:conditional-classical}

\subsubsection{Strong Pseudorandom Functions}

We first state a standard pseudorandomness consequence of the assumption that subexponentially secure one-way functions exist (\Cref{sec:crypto}). All adversaries in this subsection are
uniform randomized classical algorithms. Their oracle queries are
classical and may be adaptive.

\paragraph{Definition (Strong PRFs).}
Let $\ell\colon\mathbb N\to\mathbb N$ be polynomially bounded and
computable in polynomial time. Consider a family
$$
    \mathcal G_\lambda
    =\{g_{\lambda,s}\colon\bits^\lambda\to\bits\}
          _{s\in\bits^{\ell(\lambda)}}
$$
with a deterministic evaluation algorithm running in $\poly(\lambda)$
time on input $(1^\lambda,s,x)$. For a randomized oracle algorithm
$D$, define
$$
    \operatorname{Adv}_{\mathcal G}(D,\lambda)
    \eqdef \Big|\Pr_{s\sim\mathcal U_{\ell(\lambda)},D}
                [D^{g_{\lambda,s}}(1^\lambda)=1]
           -\Pr_{R\sim\ALL_\lambda,D}[D^R(1^\lambda)=1]\Big|,
$$
where $R$ is a uniformly random Boolean function on $\lambda$ bits.
We call $\mathcal G$ a \emph{strong pseudorandom function family} (strong PRF)
if there is a rational constant $\varepsilon\in(0,1)$ such that
every $D$ running in time at most $2^{\lambda^\varepsilon}$ satisfies
\begin{equation}\label{eq:strong-prf-security}
    \operatorname{Adv}_{\mathcal G}(D,\lambda)
    \le2^{-\lambda^\varepsilon}
\end{equation}
for every sufficiently large $\lambda$.\\

We will need the following well-known cryptographic result.

\begin{lemma}[Strong PRFs from Subexponentially Secure OWFs]
\label{lem:owf-strong-prf}
If subexponentially secure one-way functions exist, then a strong
pseudorandom function family exists. In particular, there are a rational
constant $\varepsilon\in(0,1)$ and a fixed integer $d\ge1$ such that
the family satisfies \Cref{eq:strong-prf-security} and every member
of $\mathcal G_\lambda$ has a Boolean circuit of size at most
$\lambda^d$ for all sufficiently large $\lambda$.
\end{lemma}

For completeness, we provide a proof sketch and verify the parameters in the statement of \Cref{lem:owf-strong-prf} needed in our application.

\begin{proof}[Proof Sketch]
Let $f_\lambda$ be the assumed OWF on $\lambda$ input bits, with
security exponent $\varepsilon_0>0$. We retain $\lambda$ as the
security parameter throughout.

First, we amplify the security of the assumed one-way function.
Fix a rational $0<\gamma<\varepsilon_0$ and take the
direct product of $\lambda$ independent copies of $f_\lambda$,
with total input length $\lambda^2$. The classical amplification
argument \citep[Theorem 2.3.2]{DBLP:books/cu/Goldreich2001}
embeds a challenge in each coordinate among independently generated
instances and repeats the inversion attempt. Tracking that proof,
an inverter for the direct product function of time $T$ and success $\eta\ge2^{-\lambda^\gamma}$
yields a uniform inverter for $f_\lambda$ with constant success probability in
time $\poly(T,\lambda,1/\eta)$. For $T\le2^{\lambda^\gamma}$, the resulting inverter's running time is
$2^{O(\lambda^\gamma)}<2^{\lambda^{\varepsilon_0}}$, contradicting
the assumed negligible success probability condition of $f_\lambda$. Thus every inverter for the direct product function 
running in time at most $2^{\lambda^\gamma}$ has success probability at most
$2^{-\lambda^\gamma}$, for all sufficiently large $\lambda$.

Next, we construct a subexponentially secure pseudorandom generator (strong PRG) from the amplified one-way function. In more detail, the HILL construction \citep{HILL99} 
 gives a polynomial-time pseudorandom generator
$$
    P_\lambda\colon\bits^{\ell(\lambda)}\to\bits^{2\ell(\lambda)},
    \qquad \ell(\lambda)=\poly(\lambda),
$$
with advantage at most $2^{-\lambda^\beta}$ against time
$2^{\lambda^\beta}$, for some rational $0<\beta<\gamma$.
The polynomial security loss is absorbed
by choosing $\beta>0$ sufficiently small.

Finally, we obtain a strong PRF from this pseudorandom generator. For this, we apply the GGM binary tree construction \citep{GGM86} with
depth $\lambda$ and a uniform root seed in $\bits^{\ell(\lambda)}$.
The input bits select a path, and the function outputs one bit of
the resulting leaf seed. The standard hybrid argument turns a
PRF distinguisher of time $T$ and advantage
$\delta$ into a PRG distinguisher of time
$\poly(T,\lambda)$ and advantage at least $\poly(\delta,1/\lambda,1/T)$.
Fix a rational $0<\varepsilon<\beta$. As a consequence of the preceding discussion, a violation of
\Cref{eq:strong-prf-security} would give a PRG distinguisher
with time $2^{O(\lambda^\varepsilon)}$ and advantage at least
$2^{-O(\lambda^\varepsilon)}$, contradicting the security of the pseudorandom generator
for all sufficiently large $\lambda$. Finally, each function evaluation uses
$\lambda$ generator evaluations, so fixing the key gives a circuit
of size at most $\lambda^d$ for a fixed $d$.
\end{proof}

\subsubsection{Lower Bound}

The proof of \Cref{thm:classical-lower} uses a uniformly random target,
which typically has no polynomial size circuit. To obtain a lower bound
for small circuits, we replace that target by a member of the family
from \Cref{lem:owf-strong-prf}, restricted to a smaller subcube. The
subcube is small enough that a distinguisher can evaluate the learner's
correlation exactly within the PRF security bound. Our use of a smaller
input dimension for this purpose is inspired by a similar scaling argument in the natural proofs barrier construction
\citep{RR97}.

\begin{theorem}[Conditional Classical Hardness for Polynomial Size Concepts]
\label{thm:strong-prf-separation}
Assume the existence of subexponentially secure one-way functions.
There is a fixed integer $K\ge1$ such that, for every $\rho>0$ and
$p > 1/2$, no uniform randomized classical algorithm
$\mathcal A$ has the following property: for every sufficiently
large input length $m$ and every target $f\in\mathsf{SIZE}[m^K]$
on $m$ bits, it runs in $o(2^{m/2})$ time with membership queries and
outputs a deterministic Boolean circuit $h$ satisfying
$$
    \Pr_{\mathcal A}[\corr(H,F)\ge\rho\,2^{-m/2}]\ge p,
$$
where $F(x)=(-1)^{f(x)}$ and $H(x)=(-1)^{h(x)}$.
\end{theorem}

\begin{proof}
Take $\mathcal G,d,\varepsilon$ from \Cref{lem:owf-strong-prf},
and set $K=\lceil d/\varepsilon\rceil+1$. Suppose a learner
$\mathcal A$ with the stated guarantee exists. Truncate it after
its promised  running time, and output a fixed constant
circuit if it has not produced a valid circuit by then. This
preserves its guarantee on the promised targets and bounds its
running time, query count, and output size on every oracle.

\paragraph{The Distinguisher.}
Let $\rho_\ast  \eqdef \rho/2$. In addition, at security
parameter $\lambda$, let
$$
    m\eqdef \left\lfloor\frac{\lambda^\varepsilon}{10}\right\rfloor,
    \qquad M \eqdef 2^m.
$$
Given an oracle $\mathcal{O}\colon\bits^\lambda\to\bits$, restrict it to
$$
    f'(x) \eqdef \mathcal{O}(x0^{\lambda-m}),
    \qquad F'(x) \eqdef (-1)^{f'(x)},\quad x\in\bits^m.
$$
Run $\mathcal A$ with membership access to $f'$, obtaining a
circuit $h'$ with sign representation $H'$. Next enumerate the
entire subcube and accept precisely when
\begin{equation}\label{eq:prf-correlation-test}
    \mathsf{corr}(H',F')
    \ge \rho_\ast \cdot 2^{-m/2}.
\end{equation}

\paragraph{Acceptance on a Pseudorandom Target.}
If $\mathcal{O}=g_{\lambda,s}$, then its restriction $f'$ has a circuit of
size $O(\lambda^d)=O(m^{d/\varepsilon})$. Since
$K>d/\varepsilon$, this is at most $m^K$ for all sufficiently
large $\lambda$. The learner's guarantee at threshold
$\rho \cdot 2^{-m/2}$ implies acceptance at the smaller threshold in
\Cref{eq:prf-correlation-test} with probability at least $p > 1/2$,
for every key $s$ and hence for a uniform key.

\paragraph{Acceptance on a Random Target.}
If $\mathcal{O}$ is uniformly random, $f'$ is a uniformly random Boolean
function on $m$ bits. Let $q_m=o(2^{m/2})$ bound the learner's
query complexity. For all sufficiently large
$m$ we have $\rho_\ast\sqrt M>q_m$. Apply
\Cref{thm:classical-lower} with agreement advantage
$\gamma=\rho_\ast/(2\sqrt M)>q_m/(2M)$. It follows that
$$
    \Pr[\text{accept}\mid \mathcal{O}\text{ is uniformly random}]\le\frac12.
$$

\medskip

As a consequence of the last two estimates, the distinguishing advantage is at
least the fixed positive constant $p-1/2$.

\paragraph{Running Time.}
The distinguisher runs $\mathcal{A}$ as a subroutine and makes $M$ oracle queries. Each
query takes time $\poly(\lambda)$, since we need to pad the input. The correlation can be computed in time $2^{2m} \cdot \poly(\lambda)$, since the output hypothesis $h'$ has size at most $2^{m/2}$. Thus, by computing all queries in advance,  running the learner, then computing the correlation, the total time is at most
$$
    2^m \cdot \poly(\lambda) + o(2^{m/2}) + 2^{2m} \cdot \poly(\lambda) 
    \;\le\;2^{\lambda^\varepsilon/2} \cdot  \poly(\lambda)
    \;\le\;2^{\lambda^\varepsilon},
$$
for all sufficiently large $\lambda$.  The distinguisher is a uniform randomized oracle 
algorithm within the allowed time bound, and its constant advantage
contradicts \Cref{eq:strong-prf-security}.
\end{proof}

By \Cref{eq:agreement_vs_correlation}, taking $\rho=2a_0$ and
$p=p^\star$ proves Item (ii) of \Cref{thm:classical_limitations}. The exponent $K$ depends only
on the PRF family, so the same polynomial size concept class works
for every fixed $a_0>0$ and $p^\star>1/2$.

\section{Quantum PAC Learning Advantage: Proof of \Cref{cor:padded-separation}}\label{sec:padded-separation}

Throughout this section, fix  $k\ge1$. Assume without loss of generality that $k$ is an integer. Write $\lambda_n=\lambda(n)$, and set $$r_n \eqdef 2^{k\lceil\log_2 n\rceil},$$ so that $n^k\le r_n=n^{O(1)}$. We use a standard padding argument to adjust the effective input dimension to the desired advantage. We choose targets that depend on only $m$ of their $n$ input bits, where
$$
    M=2^m=\Theta(r_n^3\lambda_n^{-2}).
$$
With this choice of parameters, $O(r_n^3)$ quantum random examples suffice for the amplified learner from \Cref{thm:quantum_learners} Item (ii) to achieve advantage $\lambda_n$ and success probability at least $1-1/n^k$. Thus, for sufficiently large $n$, it meets the required guarantee in polynomial time.

For the classical lower bound, we adapt the earlier transcript argument (\Cref{thm:classical-lower}) to padded random functions, allowing the learner's hypothesis to depend arbitrarily on all $n$ input bits. Bounding its success probability, averaged over the random target, by $1/n^k$ at advantage $\lambda_n/n^k$ suffices for the unconditional assertion of \Cref{cor:padded-separation}. In contrast, the conditional assertion concerning polynomial size concepts requires a stronger quantitative bound. We establish that the average classical success probability is bounded above by $1/(16r_n)$, leaving a noticeable gap below $1/n^k$. This gap is needed when replacing random functions by pseudorandom functions under the parameters of \Cref{cor:padded-separation}.

We first specify how ``constructivity'' lets us choose $m$, then establish the necessary query lower bound and prove the assertions of the corollary.

\paragraph{Constructive Advantage.} We say that a positive sequence $(\lambda_n)_{n\ge1}$ is \emph{constructive} if there is a deterministic algorithm that, on input $1^n$, runs in time $\poly(n)$ and outputs a nonnegative integer $s(n)$ such that, for every sufficiently large $n$,
\begin{equation}\label{eq:constructive-advantage}
    2^{-s(n)}\le\lambda_n\le4\cdot2^{-s(n)}.
\end{equation}
Thus only an approximation to the magnitude of $\lambda_n$ is required. Note that, under \Cref{eq:constructive-advantage}, the assumptions $\lambda_n2^{n/2}=n^{\omega(1)}$ and $\lambda_n=n^{-\omega(1)}$ yield
\begin{equation}\label{eq:asympt_conditions}
    s(n)=\omega(\log n),
    \qquad \frac n2-s(n)=\omega(\log n).
\end{equation}

\subsection{Refined Query Lower Bound for Padded Functions}

We will need the following second moment version of the transcript argument (\Cref{thm:classical-lower}).

\begingroup
\RenewCommandCopy{\theHlemma}{\thelemma}
\begin{lemma}[Padded Query Lower Bound]\label{lem:padded-query-lower}
Let $1\le m<n$, write $M=2^m$, and consider the target class
$$
    \mathcal D_{n,m}
    =\{f_g(x,z)=g(x):g\in\ALL_m\},
    ~\text{where} (x,z)\in\bits^m\times\bits^{n-m}.
$$
Suppose a randomized classical learner makes at most $q$ membership queries and outputs a deterministic Boolean circuit $h$ on $n$ bits. Then, for every $\gamma>q/(2M)$,
$$
    \E_{g\sim\ALL_m}\left[
        \Pr_{\mathcal A}\left[
            \Pr_{x,z}[h(x,z)=g(x)]\ge\frac12+\gamma
        \right]
    \right]
    \le\frac{M}{(2M\gamma-q)^2}.
$$
\end{lemma}
\endgroup

\begin{proof}
Choose $g$ uniformly from $\ALL_m$, fix the learner's random coins, and condition on a feasible complete transcript. The \emph{core input} of a query $y \in \{0,1\}^n$ is its $m$-bit prefix. Let $Q$ be the distinct queried core inputs, so $|Q|\le q$. As in the proof of \Cref{thm:classical-lower}, the signs $G(x)=(-1)^{g(x)}$ at unqueried inputs remain independent and uniform, even for adaptive queries. The output circuit $h$ is fixed under this conditioning. 

Define
$$
    w_x\eqdef\E_z[(-1)^{h(x,z)}]\in[-1,1] \quad \text{and}
    \quad S\eqdef\sum_{x\notin Q}  w_x \cdot G(x).
$$
These weights are used only in the analysis, allowing $h$ to depend arbitrarily on the padding bits. Since
$$
    \Pr_{x,z}[h(x,z)=g(x)]
    =\frac12+\frac1{2M}\sum_x w_x \cdot G(x)
$$
success implies
$$
    \sum_{x\in Q}w_x \cdot G(x)+S\ge2M\gamma.
$$
Each queried input $y = xz$ satisfies $w_x \cdot G(x)\le |w_x|\le1$, so
$$
    \sum_{x\in Q}w_x \cdot G(x)
    \le\sum_{x\in Q}|w_x|
    \le |Q|\le q.
$$
Consequently, on the success event,
$$
    S\ge2M\gamma-\sum_{x\in Q}w_x \cdot G(x)
    \ge2M\gamma-q>0,
$$
where the strict positivity follows from $\gamma>q/(2M)$.

Recall that we choose $g$ uniformly from $\ALL_m$, fix the learner's random coins, and condition on a feasible complete transcript. For the following calculation, all expectations, variances, and
probabilities are conditional on the transcript and coins.
Under this conditioning, the weights $w_x$ are fixed, while the
unqueried signs $G(x)$ are independent and uniform in $\{-1,1\}$.
Hence $\E[G(x)]=0$ and $\E[G(x)^2]=1$, giving
$$
    \E[S]=\sum_{x\notin Q}w_x \cdot \E[G(x)]=0.
$$
For distinct unqueried inputs $x_1,x_2$, independence gives
$\E[G(x_1)G(x_2)]=\E[G(x_1)]\E[G(x_2)]=0$. Thus all cross terms
vanish when we expand $S^2$, and
$$
    \operatorname{Var}(S)
    =\E[S^2]\\
    =\sum_{x\notin Q}w_x^2
    \le M-|Q|\le M.
$$
The first equality uses $\E[S]=0$, and the inequality uses
$|w_x|\le1$ and the fact that there are $M-|Q|$ unqueried
core inputs.

Set $a\eqdef 2M\gamma-q>0$. Since success implies $S\ge a$ and
$\E[S]=0$, the success event is contained in
$\{|S-\E[S]|\ge a\}$. Chebyshev's inequality therefore gives
$$
    \begin{aligned}
    \Pr[\text{success}]
    &\le\Pr[S\ge a]\\
    &\le\Pr[|S-\E[S]|\ge a]\\
    &\le\frac{\operatorname{Var}(S)}{a^2}\\
    &\le\frac{M}{(2M\gamma-q)^2}.
    \end{aligned}
$$
Averaging over transcripts and coins proves the lemma.
\end{proof}

\subsection{Unconditional Separation}\label{sec:uncond_sep}

In this section, we prove the unconditional part of \Cref{cor:padded-separation}.

\begin{proof}
First, we introduce some parameters and establish a few useful estimates. Let $s(n)$ be given by constructivity (\Cref{eq:constructive-advantage}), and set
$$
    \theta_n \eqdef 2^{-s(n)},
    \qquad m_n \eqdef 2s(n)+3k\lceil\log_2 n\rceil+4,
    \qquad M_n \eqdef 2^{m_n}=\frac{16r_n^3}{\theta_n^2}.
$$
The two asymptotic conditions in \Cref{eq:asympt_conditions} give $m_n=\omega(\log n)$ and
$
    n-m_n=\omega(\log n).
$
Thus $1\le m_n<n$ eventually. The parameter choice has three useful consequences:
\begin{equation}\label{eq:padded-parameters}
    M_n\theta_n^2=16r_n^3,
    \qquad M_n\lambda_n^2\le256r_n^3,
    \qquad \frac{M_n\theta_n}{r_n}=\frac{16r_n^2}{\theta_n}=n^{\omega(1)}.
\end{equation}

Next, define $\mathcal C_n \eqdef \mathcal D_{n,m_n}$ (as in \Cref{lem:padded-query-lower}) for all sufficiently large $n$. At the finitely many remaining lengths, take the singleton containing the constant zero function and let the quantum learner output that function.

\paragraph{Quantum Learner.} First, we observe that quantum random examples for $f_g$ supply quantum random examples for $g$. In more detail, compute $m_n$ using the algorithm for $s(n)$. After permuting registers, a quantum random example for $f_g(x,z)=g(x)$ factors as
$$
    |\mathrm{QEX}_{f_g}\rangle
    =\left(2^{-m_n/2}\sum_x|x\rangle|g(x)\rangle\right)
      \otimes|+\rangle^{\otimes(n-m_n)},
$$
where we have used that, for $b=n-m_n$,
$$
    |+\rangle^{\otimes b}
    =\left(\frac{|0\rangle+|1\rangle}{\sqrt2}\right)^{\otimes b}
    =2^{-b/2}\sum_{z\in\bits^b}|z\rangle.
$$
Discard the padding registers and apply \Cref{thm:quantum_learners} Item~(ii) to the resulting independent examples for $g$, with $\eta_n \eqdef 1/(Lr_n^3)$, where $L\ge\max\{1,256/a_1^2\}$ is a fixed integer. Its condition $C2^{-m_n}\le\eta_n<1/2$ holds eventually because $\eta_nM_n=16/(L\theta_n^2)\to\infty$. With probability at least $1-\eta_n\ge1-1/n^k$, the output $h_0$ satisfies
$$
    \Pr_x[h_0(x)=g(x)]
    \ge\frac12+a_1\sqrt{\frac{Lr_n^3}{M_n}}
    \ge\frac12+\lambda_n,
$$
where the last inequality follows from $\lambda_n\sqrt{M_n}\le16r_n^{3/2}$ and $a_1\sqrt L\ge16$. Output $h(x,z)=h_0(x)$, preserving agreement. Computing $s(n)$ takes polynomial time, and processing the $O(r_n^3)$ examples and constructing the output circuit costs
$\poly(n)$ time in total due to our choice of $\eta_n=1/(Lr_n^3)$.

\paragraph{Classical Lower Bound.}
Fix a classical polynomial time learner $\mathcal B$, with polynomial query bound $q(n)$. By \Cref{eq:padded-parameters}, $q(n)\le M_n\theta_n/r_n$ eventually. Apply \Cref{lem:padded-query-lower} at advantage $\theta_n/r_n$ to obtain
\begin{equation}\label{eq:padded-random-gap}
    \E_{g\sim\ALL_{m_n}}\left[
        \Pr_{\mathcal B}\left[
            \Pr_{x,z}[h(x,z)=g(x)]\ge\frac12+\frac{\theta_n}{r_n}
        \right]
    \right]
    \le\frac{r_n^2}{M_n\theta_n^2}=\frac1{16r_n}.
\end{equation}
Since $\theta_n/r_n\le\lambda_n/n^k$, the same bound applies at advantage $\lambda_n/n^k$. Averaging provides a target function with success probability at most $1/(16r_n)$, and hence at most $1/n^k$, at every sufficiently large length.
\end{proof}

\subsection{Conditional Separation for $\mathfrak{C} \subseteq \mathsf{SIZE}[\poly]$}

In this section, we prove the conditional part of \Cref{cor:padded-separation}. For this, we replace the random (core) functions by restrictions of strong PRFs. We enlarge the security parameter so that exhaustive agreement testing on all $2^n$ inputs remains within the PRF security bound, while the resulting concepts still have polynomial size.

\begin{proof}
Define $s(n)$, $\theta_n$, $r_n$, and $m_n$ as in  the unconditional construction (\Cref{sec:uncond_sep}), and let $\mathcal G,\varepsilon,d$ be as in \Cref{lem:owf-strong-prf}. Choose a fixed integer $A$ with $A > 1/\varepsilon$ and use security parameter $t_n=n^A$. Replace the concept class above, at sufficiently large input lengths, by
$$
    \mathcal C_n^{\mathrm{PRF}}
    =\left\{(x,z)\longmapsto
        g_{t_n,\sigma}(0^{t_n-m_n}\concat x):
        \sigma\in\bits^{\ell(t_n)}\right\},
$$
where $\ell(t_n)$ is the PRF key length, $|x| = m_n$, and $|x| + |z| = n$. Hardwiring the key and zero prefix gives circuits of size at most $t_n^d=n^{Ad} = \poly(n)$. 

The quantum upper bound argument applies unchanged. This follows from our choice of parameters (as in the proof given in  \Cref{sec:uncond_sep}) and the fact that the quantum learner from \Cref{thm:quantum_learners} Item (ii) applies to every target Boolean function. 

Next, we obtain a conditional lower bound against classical learners. This part depends on the security of the strong PRF, which in turn relies on the assumption that subexponentially secure OWFs exist.

\paragraph{Distinguisher Construction and Advantage.}
Fix a uniform randomized polynomial time learner $\mathcal B$, and suppose that at infinitely many lengths $n$ it succeeds with probability greater than $1/n^k$ on every target in $\mathcal C_n^{\mathrm{PRF}}$ at advantage $\lambda_n/n^k$. Let $\mathcal{I}$ denote the set of such input lengths. 

The distinguisher computes as follows. For an oracle $\mathcal O\colon\bits^{t_n}\to\bits$, simulate $\mathcal B$ on
$$
    f^{\mathcal O}(x,z)=\mathcal O(0^{t_n-m_n}\concat x).
$$
After the learner outputs $h$, accept exactly when
$$
    \Pr_{x,z}[h(x,z)=f^{\mathcal O}(x,z)]
    \ge\frac12+\frac{\theta_n}{r_n}.
$$

For a random oracle, the selected restriction is uniform in $\ALL_{m_n}$, so \Cref{eq:padded-random-gap} bounds the acceptance probability by $1/(16r_n)$. On the other hand, for an input length $n \in \mathcal{I}$, every PRF key gives acceptance probability greater than $1/n^k\ge1/r_n$, because success at advantage $\lambda_n/n^k$ implies acceptance at the smaller threshold $\theta_n/r_n$. Thus this test distinguishes the two oracle distributions with advantage at least $15/(16r_n)$. 

\paragraph{Distinguisher Uniformity and Running Time.}
To implement the test by a single uniform distinguisher $D$, on input $1^t$ reject unless $t=n^A$ for an integer $n$ above a sufficiently large fixed threshold, and then compute $s(n)$, $r_n$, and $m_n$. Clock $\mathcal B$ at a sufficiently large polynomial bound on every oracle, with a fixed circuit hypothesis as fallback if the clock expires. This does not change its behavior on the PRF targets, and ensures that the random oracle simulation also has polynomial running time.

Evaluate $h$ and $f^{\mathcal O}$ on all $2^n$ inputs to test their agreement exactly. Including the simulation, circuit evaluations, and oracle queries of length $t_n=\poly(n)$, the running time is
$$
    2^n \cdot \poly(n,t_n)\le 2^{n^{A\varepsilon}}=2^{t_n^\varepsilon}
$$
for all sufficiently large $n$, by the choice of $A >1/\varepsilon$ and $t_n = n^A$.\\ 

Since $r_n=n^{O(1)}$, the distinguishing advantage $15/(16r_n)$ exceeds $2^{-t_n^\varepsilon}$ for every sufficiently large $n$. Together with the running time bound above, this violates \Cref{eq:strong-prf-security} at infinitely many security lengths $t_n$. This contradicts the security of the PRF family. Hence every sufficiently large $n$ has a target in $\mathcal C_n^{\mathrm{PRF}}$ on which $\mathcal B$ succeeds at advantage $\lambda_n/n^k$ with probability at most $1/n^k$.
\end{proof}

\small

\bibliographystyle{alpha}
\bibliography{references}

\normalsize

\appendix

\section{Fourier-Based Quantum Learner for $\mathcal{C} = \ALL$ with $p^\star \leq 1/2$}\label{sec:fourier_sampling}

In this section, we review the Fourier-based quantum learner for $\mathcal{C} = \ALL$ described in \citep[Appendix A]{AGGOS21}. This learner achieves correlation $\geq \gamma \cdot 2^{-n/2}$ and its success probability $p^\star  \to  1/2$ from below as $\gamma \to 0$.

\paragraph{The Key Idea.}
 Recall that, for a sign-valued function $F\colon \bits^n\to\pmone$, its Fourier expansion is\footnote{The necessary background in Fourier analysis of Boolean functions can be found in a standard textbook such as \citep{o2014analysis}.}
\[
    F(x)=\sum_{S\subseteq[n]}\widehat F(S)\chi_S(x),
    \qquad
    \chi_S(x)=(-1)^{S\cdot x},
    \qquad
    \widehat F(S) = \E_{x \sim \Un}[\chi_S(x) \cdot F(x)],
\]
where $S \cdot x$ denotes  $\sum_{i \in S} x_i~(\mathsf{mod}\;2)$. 
Crucially for us, $\widehat F(S)$ is the \emph{correlation} between $F$ and the parity function  $\chi_S$.

Parseval's identity gives
\[
    \sum_S \widehat F(S)^2=1,
\]
so $\{\widehat F(S)^2\}_S$ is a probability distribution. We use $S\sim\widehat F^2$ to denote a set $S$ sampled from this distribution. The main observation is that with constant probability we obtain a nontrivial correlation.

\begin{claim}
For every $0\le \gamma\le1$ and every $F\colon\bits^n\to\pmone$,
\[
    \Pr_{S\sim \widehat F^2}\left[|\widehat F(S)|\ge \gamma \cdot 2^{-n/2}\right]
    \ge 1-\gamma^2 .
\]
\end{claim}

\begin{proof}
Let $N = 2^n$, and consider the set 
$$
    B=\left\{S:|\widehat F(S)|<\frac{\gamma}{\sqrt N}\right\}.
$$
Since Fourier sampling assigns probability $\widehat F(S)^2$ to $S$,
\begin{equation}\nonumber
    \Pr_{S\sim\widehat F^2}[S\in B]
    \; = \;
    \sum_{S\in B}\widehat F(S)^2
    \; \le \;
    |B|\frac{\gamma^2}{N}
    \;\le\; \gamma^2.
\end{equation}
Taking the complementary event proves the claim.
\end{proof}

Consequently, a Fourier-sampled parity $\chi_S$ or its negation has correlation
$\Omega(2^{-n/2})$ with $F$ with constant probability.  The obstruction is the \emph{sign}: Fourier sampling reveals $S$, but not whether $\chi_S$ or $-\chi_S$ is positively correlated with $F$.  If the learner simply guesses the sign, its orientation succeeds with probability $1/2$ conditional on having sampled a sufficiently large coefficient. For this reason, the learner achieves constant success probability, but this probability is not larger than $1/2$.

\paragraph{The Standard Fourier Quantum Sampler.} It remains to explain how to sample from the distribution $\widehat F^2$. We use the standard Fourier sampling procedure, implemented using access to uniform quantum examples. Recall that
$$
|\mathrm{QEX}_f\rangle
=
\frac{1}{\sqrt N}\sum_{x\in\bits^n}|x\rangle|f(x)\rangle,
\qquad
F(x)=(-1)^{f(x)}.
$$

The procedure consists of three steps. First, we convert the quantum example into a phase state. Applying a Hadamard gate to the label qubit gives
$$
\begin{aligned}
(I^{\otimes n}\otimes\Had)|\mathrm{QEX}_f\rangle
&=
\frac{1}{\sqrt N}
\sum_x
|x\rangle
\Had|f(x)\rangle \\
&=
\frac{1}{\sqrt{2N}}
\sum_x
|x\rangle
\bigl(
|0\rangle+(-1)^{f(x)}|1\rangle
\bigr) \\
&=
\frac{1}{\sqrt2}
\left(
\frac{1}{\sqrt N}\sum_x|x\rangle
\right)|0\rangle
+
\frac{1}{\sqrt2}
\left(
\frac{1}{\sqrt N}\sum_xF(x)|x\rangle
\right)|1\rangle.
\end{aligned}
$$
Define the phase state
$$
|F\rangle
\eqdef
\frac{1}{\sqrt N}\sum_xF(x)|x\rangle.
$$
The state above can then be written as
$$
\frac{1}{\sqrt2}
\left(
\frac{1}{\sqrt N}\sum_x|x\rangle
\right)|0\rangle
+
\frac{1}{\sqrt2}|F\rangle|1\rangle.
$$

Second, we measure the label qubit in the computational basis. Since the two displayed branches each have squared norm $1/2$, the outcome $1$ occurs with probability exactly $1/2$. Conditional on this outcome, the input register is left in the state $|F\rangle$. If the outcome is $0$, we discard the input register and repeat the procedure using a fresh quantum example. 

Third, after obtaining $|F\rangle$, we apply $\Had^{\otimes n}$ to the input register. Since
$$
\Had^{\otimes n}|x\rangle
=
\frac{1}{\sqrt N}
\sum_{S\in\bits^n}
(-1)^{S\cdot x}|S\rangle,
$$
we obtain
$$
\begin{aligned}
\Had^{\otimes n}|F\rangle
&=
\frac{1}{N}
\sum_{S\in\bits^n}
\sum_{x\in\bits^n}
F(x)(-1)^{S\cdot x}|S\rangle \\
&=
\sum_{S\in\bits^n}
\widehat F(S)|S\rangle,
\end{aligned}
$$
where
$$
\widehat F(S)
=
\frac{1}{N}
\sum_{x\in\bits^n}
F(x)(-1)^{S\cdot x}.
$$
A measurement of the input register in the computational basis therefore returns $S$ with probability
$$
|\widehat F(S)|^2
=
\widehat F(S)^2.
$$
Hence the measurement outcome after this step is distributed exactly according to $\widehat F^2$, as desired.

\section{Orienting a Weak Hypothesis is Hard}
\label{sec:orientationHardness}

Let $N=2^n$ and let $H\colon\bits^n\to\pmone$ be a known hypothesis
for an unknown target $F\colon\bits^n\to\pmone$. Under the promise
$$|\E_x[F(x)H(x)]|\ge c/\sqrt N$$ for fixed $c>0$ and uniform $x$,
\emph{orienting} $H$ means choosing $H$ or $-H$ to obtain \emph{positive correlation}.

Consider a string $z \in \{0,1\}^N$ indexed by $x \in \{0,1\}^n$, where $$z(x) \eqdef (1-F(x)H(x))/2.$$ Since $H$ is known, query access
to $F$ and to $z$ are equivalent.
Since $\E_x[F(x)H(x)]=1-2|z|_1/N$, where $|z|_1\eqdef\sum_x z(x)$,
orientation is the ``gap majority problem'', i.e., we need to decide whether
$$
    |z|_1\le N/2-c\sqrt N/2
    \qquad\text{or}\qquad
    |z|_1\ge N/2+c\sqrt N/2.
$$

Nayak and Wu~\citep[Corollary~1.2]{NayakWu99} show that distinguishing
the two boundary weights  requires
$$
    \Omega(\sqrt N)=\Omega(2^{n/2})
$$
\emph{quantum membership queries}\footnote{A quantum query is defined by a unitary map $O_z$ acting on $n + 1$ qubits whose action on basis states is defined as  $O_z \colon |x,b\rangle \rightarrow |x,b\oplus z(x)\rangle$, for every $x \in \{0,1\}^n$ and $b\in\bits$. A $q$-query quantum algorithm makes at most $q$ such calls. Computation between
calls is unrestricted.} for any fixed success probability greater than $1/2$ on every
promised input. 

Consequently, any quantum algorithm that orients an arbitrary known hypothesis $H$
with fixed success probability greater than $1/2$ for every promised target $F$ requires $\Omega(2^{n/2})$
membership queries to the unknown target $F$ in the worst case. The quantum learner in \Cref{sec:tiltedSamplerAlgorithm} avoids this obstacle. Its measurement distribution favors positively correlated hypotheses,
without first fixing a candidate and then determining its sign.

\section{Limits of Parity Functions and of Pairwise Independence}
\label{sec:whyParitiesDoNotSuffice}

Both the classical learner in \Cref{prop:zero-query-construction}
and the quantum learner in \Cref{thm:quantum_correctness} use the
quadratic Boolean family from \Cref{lem:prop_family}. We examine
what happens when this family is replaced by affine parities.
The resulting family is efficiently computable, pairwise
independent, and admits a global sign flip. Nevertheless, as explained below, this
replacement fails to preserve the success guarantee of either
learner.

\paragraph{The Classical Learner: Pairwise Independence Does Not
Guarantee Constant Success Probability.}
The classical learner in \Cref{prop:zero-query-construction}
samples a uniform seed and outputs the corresponding hypothesis.
With the quadratic family, it achieves agreement at least
$1/2+1/(4\sqrt N)$ with probability at least $3/32$ for every
target, where $N=2^n$.

Consider instead the family of affine parities
$$
    h_{u,b}(x)\eqdef(u\cdot x)\oplus b,
    \qquad \chi_u(x)\eqdef(-1)^{u\cdot x},
    \qquad H_{u,b}(x)\eqdef(-1)^b\chi_u(x) = (-1)^{h_{u,b}(x)},
$$
where $u\in\bits^n$ and $b\in\bits$ are independent and uniform.
Write $\mathsf U$ for this seed distribution.
The signs $H_{u,b}(x)$ are unbiased and pairwise independent. Changing $b$ to $b\oplus1$ also gives a global sign flip.

Fix a parity target $f(x)=s\cdot x$, with sign representation
$F=\chi_s$. Orthogonality of the Fourier characters gives
$$
    Z_{u,b}\eqdef\sqrt N\,\corr(H_{u,b},F)
    =
    \begin{cases}
        \sqrt N,&(u,b)=(s,0),\\
        -\sqrt N,&(u,b)=(s,1),\\
        0,&u\ne s.
    \end{cases}
$$
Consequently,
$$
    \E_{\mathsf U}[Z]=0,
    \qquad
    \E_{\mathsf U}[Z^2]=1,
    \qquad
    \E_{\mathsf U}[Z^4]=N,
$$
but
$$
    \Pr_{\mathsf U}[|Z|\ge1/2]=\frac1N.
$$
In particular, the success probability of the classical learner
using this family is exactly
$$
\begin{aligned}
    &\Pr_{(u,b)\sim\mathsf U}\!\left[
        \Pr_{x\sim\mathcal U_n}[h_{u,b}(x)=f(x)]
        \ge\frac12+\frac1{4\sqrt N}
    \right]\\
    &\qquad=\Pr_{\mathsf U}[Z\ge1/2]
    =\frac1{2N}.
\end{aligned}
$$

This example identifies the role of the fourth moment bound in
the analysis of \Cref{prop:zero-query-construction}.
Applying the Paley-Zygmund inequality
(\Cref{lem:PZ-Inequality}) to $X=Z^2$, using $\E_{\mathsf U}[Z^2]=1$
and $\lambda=1/4$, gives
$$
    \Pr_{\mathsf U}[|Z|\ge1/2]
    \ge\frac{9}{16\E_{\mathsf U}[Z^4]}.
$$
The global sign flip makes the distribution of $Z$ symmetric,
and hence
$$
    \Pr_{\mathsf U}[Z\ge1/2]
    =\frac12\Pr_{\mathsf U}[|Z|\ge1/2]
    \ge\frac{9}{32\E_{\mathsf U}[Z^4]}.
$$
For the quadratic family, \Cref{lem:fourWiseMomentBounds} gives
$\E_{\mathsf U}[Z^4]\le3$, yielding the bound $3/32$ used in
\Cref{eq:prob_bounds} and in the classical learner's analysis.
For affine parities, the same argument gives only $9/(32N)$,
consistent with the exact success probability $1/(2N)$.
Thus pairwise independence does not suffice to guarantee
$\Omega(1)$ success probability for this classical sampling
procedure.

\paragraph{The Quantum Learner: Affine Parities Do Not Preserve
the Success Guarantee Above $1/2$.}
The state preparation and family transform in
\Cref{sec:tiltedSamplerAlgorithm} remain well defined for affine
parities. With seeds encoded as $(u,b)$, the feature map
$$
    J(\perp)=0^{n+1},
    \qquad
    J(x)=(x,1)
$$
is injective, has an efficiently computable inverse on its image,
and satisfies $H_{u,b}(x)=(-1)^{(u,b)\cdot J(x)}$.
Since there are $2N$ seeds, the quantum measurement distribution
for an arbitrary target $F$ is
$$
    \mathsf P_F(u,b)=\frac{(1+Z_{u,b})^2}{4N},
    \qquad
    Z_{u,b}=(-1)^b\sqrt N\,\widehat F(u),
$$
where $\widehat F(u)=\E_{x\sim\mathcal U_n}[F(x)\chi_u(x)]$.

For the parity target $F=\chi_s$ considered above, the only
output with positive correlation is $H_{s,0}=F$. Substituting
the three possible values of $Z$ gives
$$
\begin{aligned}
    \mathsf P_F(s,0)
    &=\frac14+\frac1{2\sqrt N}+\frac1{4N},\\
    \mathsf P_F(s,1)
    &=\frac14-\frac1{2\sqrt N}+\frac1{4N},\\
    \sum_{u\ne s,\,b}\mathsf P_F(u,b)
    &=\frac12-\frac1{2N}.
\end{aligned}
$$
Thus the probability of meeting the agreement threshold in
\Cref{thm:quantum_correctness} is exactly
$$
    \frac14+\frac1{2\sqrt N}+\frac1{4N},
$$
which tends to $1/4$. In other words, while the quantum procedure admits 
$\Omega(1)$ success probability on this target, it fails to
achieve success probability $> 1/2$.

\end{document}